\documentclass[12pt, draftclsnofoot, onecolumn]{IEEEtran}
\usepackage{amsfonts}
\usepackage{amsmath}
\usepackage{multirow}
\usepackage{graphicx}
\usepackage{amsfonts}
\usepackage{amsmath,epsfig}
\usepackage{stfloats}
\usepackage{amssymb}
\usepackage{url}
\usepackage{bm}
\usepackage{cite}
\usepackage{amsmath}
\usepackage{amssymb}
\usepackage{latexsym}
\usepackage{multirow}
\usepackage{epsfig}
\usepackage{graphics}
\usepackage{mathrsfs}
\usepackage{algorithmic}
\usepackage{algorithm}
\usepackage{subfigure}
\usepackage[all]{xy}

\usepackage{pifont}
\usepackage{bbding}
\usepackage{stmaryrd}
\usepackage{amssymb}
\usepackage{amsfonts}
\usepackage{epic}
\usepackage{stfloats}
\usepackage{latexsym}
\usepackage{epstopdf}
\usepackage{epic}
\usepackage{bm}
\usepackage{xcolor}
\usepackage{mathrsfs}
\usepackage{pifont}
\usepackage{bbding}
\usepackage{amsmath,epsfig}
\usepackage{stmaryrd}
\usepackage{amssymb}
\usepackage{amsfonts}
\usepackage{epic}
\usepackage{graphicx}
\usepackage{subfigure}

\usepackage{enumerate}

\usepackage{stfloats}
\usepackage{latexsym}
\usepackage{epstopdf}
\usepackage{epic}
\usepackage{multirow}
\usepackage{stfloats}
\usepackage{bm}
\usepackage{amsmath}
\usepackage{amssymb}
\usepackage{amsthm}

\usepackage{booktabs}
\usepackage{color}
\usepackage{cases}
\usepackage{array}
\usepackage{makecell}
\usepackage{times}

\usepackage{url}

\usepackage{threeparttable}
\theoremstyle{definition}

\newcommand{\mv}[1]{\mbox{\boldmath{$ #1 $}}}
\newcommand{\Q}{\bm Q}
\newcommand{\A}{\bm A}

\newcommand{\II}{\mathcal{I}}

\newcommand{\Ss}{\bm S}

\newcommand{\V}{\bm V}

\newcommand{\W}{\bm W}

\newcommand{\F}{\bm F}

\newcommand{\ttheta}{\mathbf \Theta}

\newcommand{\N}{\mathcal{N}}

\newcommand{\q}{\bm q}

\newcommand{\uuu}{\bm u}

\newcommand{\vvv}{\bm v}

\newcommand{\g}{\bm g}

\newcommand{\K}{\mathcal{K}}

\newcommand{\E}{\mathcal{E}}

\graphicspath{{./figs/}}

\begin{document}
\title{Intelligent Reflecting Surface Aided Wireless Energy and Information Transmission:  An Overview}

\author{\IEEEauthorblockN{Qingqing Wu, \emph{Member, IEEE}, Xinrong Guan, and Rui Zhang,  \emph{Fellow, IEEE}
\thanks{
Q. Wu is with the State Key Laboratory of  Internet of Things for Smart City, University of Macau, Macau, 999078, China (email: qingqingwu@um.edu.mo).
X. Guan is with the College of Communications Engineering, Army Engineering University of PLA, Nanjing, 210007, China (e-mail: guanxr@ieee.org).
 R. Zhang is  with the Department of Electrical and Computer Engineering, National University of Singapore, Singapore 117583 (e-mail: elezhang@nus.edu.sg).
 } }
 \IEEEspecialpapernotice{(Invited Paper)}  }

\maketitle

\vspace{-0.5cm}
\begin{abstract}
Intelligent reflecting surface (IRS) is a promising technology for achieving spectrum and energy efficient wireless networks cost-effectively. Most existing works on IRS have focused on exploiting IRS to enhance the performance of wireless communication or wireless information transmission (WIT), while its potential for boosting the efficiency of radio-frequency (RF) wireless energy transmission (WET) still remains largely open. Although IRS-aided WET shares similar characteristics with IRS-aided WIT, they differ fundamentally  in terms of design objective, receiver architecture, practical constraints, and so on.  In this paper, we provide a tutorial overview on how to efficiently design  IRS-aided WET systems as well as IRS-aided systems with both WIT and WET, namely IRS-aided simultaneous  wireless information and power transfer (SWIPT) and IRS-aided wireless powered communication network (WPCN), from a communication and signal processing perspective. In particular, we present state-of-the-art solutions to tackle the unique challenges in operating these systems, such as IRS passive reflection optimization, channel estimation and deployment. In addition, we propose new solution approaches and point out important directions for future research and investigation.
\end{abstract}

\newpage
\section{Introduction}

\subsection{Overview of Wireless Power Transfer (WPT)}
The number of Internet-of-Things (IoT) devices  is expected  to reach an unprecedentedly high figure of  30.9 billions globally by 2025, as compared to the currently estimated   13.8 billions  in 2021 \cite{Iotnumber2021}. This is mainly driven by their increasingly immersed applications in different sectors of our society, such as  healthcare, manufacturing, transportation, smart home, etc. To  unleash the full potential of such an explosion of IoT devices in the future, cost-effective  solutions are needed to provide them with not only reliable communication over the air,   but also perpetual energy supply. To this end,  radio-frequency (RF) transmission enabled far-field wireless energy transmission (WET)/wireless power transfer (WPT) has emerged as a practically appealing technology  for  powering  IoT devices without wire \cite{zhangrui13_mimo,zeng2017communications,clerckx2018fundamentals,clerckx2021wireless}.
Compared with the conventional battery or inductive/magnetic resonant coupling based near-field wireless  charging solution,  RF WPT eliminates the hassle of  battery replacement and also significantly extends the near-field wireless charging distance.  RF WPT  has thus gained an upsurge of interest from both academia and industry in the last decade. For example, China's leading information technology company Xiaomi has recently announced their self-developed ``Mi Air Charge'', which was able to deliver up to 5 watt (W) power over the air simultaneously to multiple devices within a distance  of several meters \cite{miair2021}. Other companies dedicated to commercializing wireless charging solutions based on  RF WPT  include Powercast, TransferFi, Energous, etc.

Despite its convenience and high potential, RF WPT usually operates with low energy efficiency due to  wireless channel impairments such as  path loss, shadowing, and multi-path fading, especially when the distance from the energy transmitter (ET) to the energy receiver (ER) is long. Although increasing the radiation power at the ET can increase the received power at the ER, it does not help improve the end-to-end efficiency and may be even  infeasible due to practical constraints on radio signal's  radiation/absorption power for safety consideration \cite{zhangrui13_mimo}. Moreover, ERs in general require much higher received signal power (e.g. several tens of dB more) as compared to the information receivers in conventional wireless communication or wireless information transmission (WIT) systems \cite{zhangrui13_mimo}.
As such, how to combat wireless channels for improving the  WPT  efficiency is a crucial but challenging  problem to solve.
On the other hand, to reduce the cost of implementing WPT in practice, a promising  paradigm is to integrate WPT to existing WIT systems, namely wireless information and power transmission (WIPT) such that the information access points (APs) and communication spectrum can be also used to enable WPT  \cite{zhangrui13_mimo,clerckx2018fundamentals}. Depending on application scenarios, WIPT can be further divided into two fundamental models, namely simultaneous wireless  information and power transfer (SWIPT) and wireless powered communication network (WPCN).  To be specific, information and power in  SWIPT are transmitted concurrently from the same AP using the same RF waveform to wireless devices (e.g., wireless powered actuators)  in the downlink. As such,  the performances of WIT and WPT depend on the transmit signal waveform and exhibit  a fundamental rate-energy tradeoff.  By contrast, in WPCN,  wireless devices (e.g., wireless powered sensors)  harvest energy from signals sent by the AP in the downlink, and use their harvested energy to transmit information back to the AP  in the uplink.  In both above WIPT systems,  the low WPT efficiency  is usually the main limiting factor to their achievable performance in practice.

To maximize the end-to-end efficiency of WPT (or equivalently, attain the maximal output direct current (DC) power at the ER without increasing the transmit power at the ET), great  research effort has been made in the past to design efficient RF hardware components such as  circuits, antennas, rectifiers, etc. However, this approach cannot adapt to wireless channel variations in space, time and frequency.
As such, a new approach was proposed in wireless communication research community by applying advanced communication and signal processing techniques to WPT, which has gained increasing attention in the last decade \cite{zeng2017communications}. Several techniques in WIT have been re-investigated for WPT, such as    multiple-antenna energy beamforming, energy feedback, energy waveform design, multipoint cooperative transmission, mobile chargers, etc \cite{zeng2017communications}. Moreover,  to improve the performance of  WIPT systems, sophisticated techniques have also been proposed for SWIPT and WCPN, such as  joint information and energy beamforming design, joint waveform and beamforming design, joint communication and energy scheduling, and so on \cite{clerckx2018fundamentals,clerckx2021wireless}. Despite significantly improving  the WPT/WIPT performance over the traditional hardware-driven  approach, these communication and signal processing techniques may incur high cost in practice, but still not be able  to overcome very severe power loss due to long transmission distance and wireless channel impairments. For example, sharply focused energy beams require a large number of active transmit antennas and RF chains (144 antennas for the aforementioned ``Mi Air Charge''), which are not only  bulky and costly but power-hungry  as well \cite{wu2016overview,miair2021,buzzi2016survey}. In addition, densely deploying APs for multipoint transmission  based WPT incurs excessive energy consumption and deployment/maintenance cost \cite{wu2016energy,zhang2016fundamental}. Although leveraging the maneuverability of ground/aerial vehicle-mounted mobile chargers  is able to shorten the WPT service distance with enhanced efficiency,  trajectories of  ground vehicles are largely subjected to complex terrestrial terrains, thus with less flexibility as compared to aerial vehicles; however,  the endurance of aerial vehicles (such as unmanned aerial vehicle (UAV) or drone)  is typically constrained by their limited on-board battery. Furthermore, due to  practical speed constraints, mobile chargers need sufficient  time to move close to wireless devices to charge them, and thus may not be able to guarantee the energy charging requirement in real time \cite{wu2019UAVtdmaga,zeng2019accessing}. Considering the above issues, it is still imperative to develop new and innovative technologies to  boost the efficiency of WPT as well as the performance of WIPT systems  at a sustainable  cost.


\subsection{IRS-aided WIT and WPT}

Recently, intelligent reflecting surface (IRS) has been proposed as a promising technique to achieve high spectral and energy efficiency for future wireless networks cost-effectively \cite{wu2018IRS,JR:wu2018IRS}.  Specifically, IRS  consists of a large number of low-cost reflecting elements, each of which can be adjusted in real time to tune the amplitude and/or phase of   impinging signals, without the need of costly and power-hungry RF chains. With densely deployed IRSs in wireless networks and via smart coordination of their reflections, wireless propagation channels can be proactively  reconfigured to enhance the communication performance. As such, integrating IRSs into wireless communications provides a novel  approach, which is drastically different from the conventional one addressing  the transceivers/endpoints design only, while regarding the wireless propagation environment as a random and uncontrollable medium \cite{basar2019wireless,JR:wu2018IRS}. Furthermore, IRSs have other appealing features, such as low profile, lightweight, conformal geometry, which make them easy to be installed on the surface of environment objects (e.g. buildings, billboards, and vehicles). IRSs can also be designed to operate in different wireless systems such as cellular and WiFi, with their existing access points and user terminals, thus featuring both high compatibility and scalability \cite{JR:wu2019IRSmaga,di2020smart_JSAC}.  Due to the above advantages, IRS-aided wireless communications have been thoroughly  studied for various system setups, including single-user/multiuser/multicell systems, physical layer security,  non-orthogonal multiple access (NOMA), millimeter wave (mmWave) communications, and so on  \cite{JR:wu2019IRSmaga,JR:wu2018IRS,cui2019secure,guan2019intelligent,chen2019intelligent,fu2019intelligent,yang2019intelligent,
han2018large,huangachievable,yan2019passive,wang2019intelligent,jamali2019intelligent,9206080,dongfang2019,xu2020resource,9264659,rajatheva2020white}.

\begin{figure}[!t]
	\centering
	\includegraphics[width=1\textwidth]{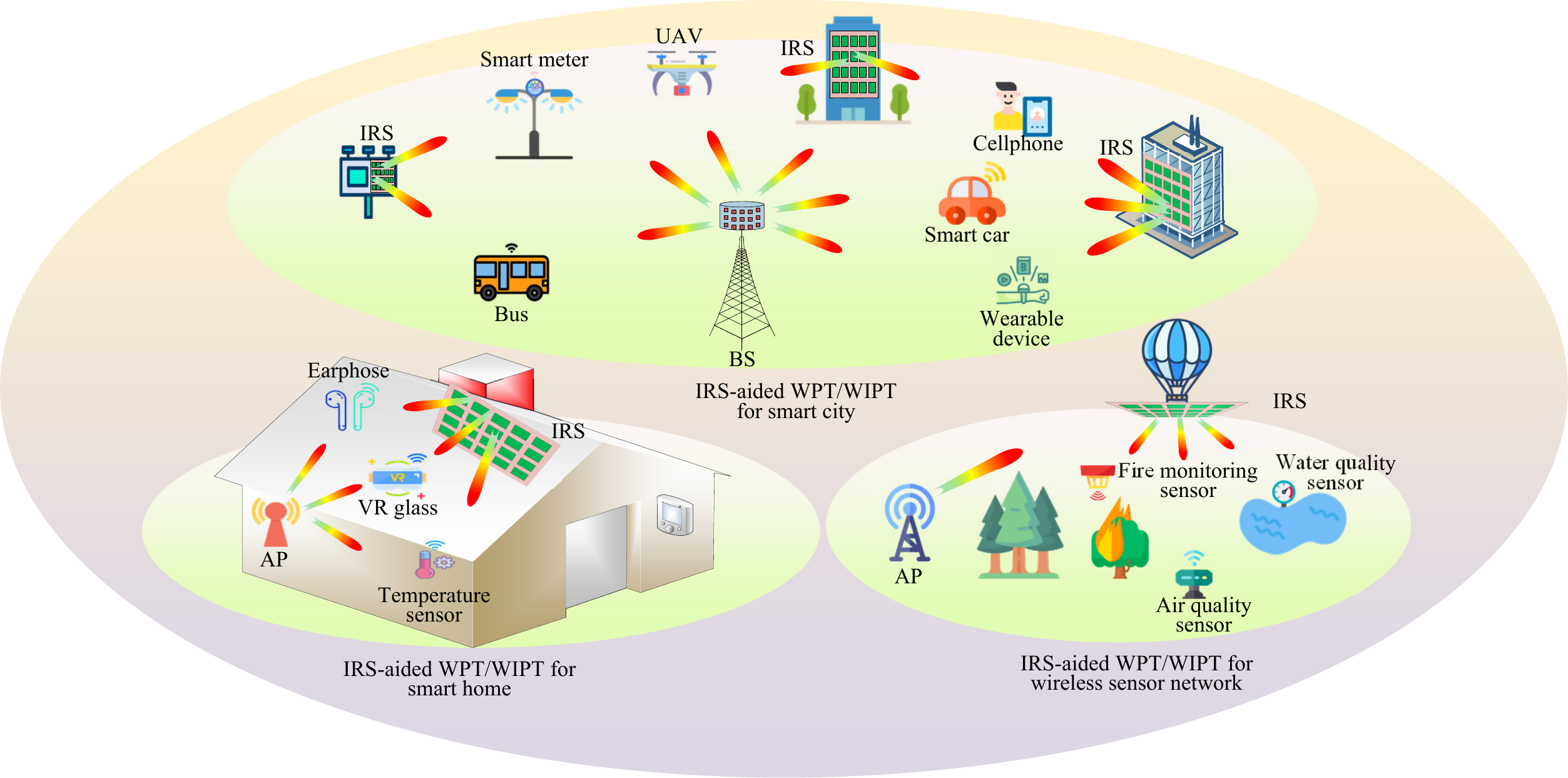}\vspace{-0.05cm} 
	\caption{IRS-aided WPT/WIPT for various IoT applications. } \label{overview}\vspace{-0.2cm}
\end{figure}

Most of the existing works on IRS considered  IRS-aided communication or WIT, while its channel reconfiguration ability and promising passive beamforming gain are also highly desirable for enhancing the efficiency of  WPT as we well as the performance of WIPT systems. For example, as shown in Fig. \ref{overview}, by leveraging smart reflections over their large aperture,  IRSs can create sharp passive energy beams to compensate the high signal attenuation over long distance and thereby establish enhanced wireless charging zones for wireless devices in hot-spot areas.  This is practically useful  for significantly  extending the coverage of WPT.  Moreover, for conventional WPT systems with severely degraded  efficiency due to blockage, deploying a sizable IRS that has line-of-sight (LoS) links with the AP and users can guide  energy signals towards the users by  bypassing obstacles. This would help ensure  WPT performance even in non-LoS (NLoS) environments. As such, IRS-aided WPT is a promising solution to supply  wireless power for   IoT devices of all kinds to reduce their use of battery.  Motivated by this, a preliminary IRS prototype with  1-bit programmable reflecting elements of size  16$\times$16 has been recently developed  \cite{tran2019novel,amri2020programmable}, which shows the effectiveness  of IRS for enhancing the efficiency of WPT under practical  setups.

However, different from  traditional WPT systems comprising active components only, IRS-aided WPT systems consist of both active (ETs and ERs) and passive (IRS) nodes. This thus gives rise to new challenges in designing their joint operation and in particular new and important issues brought by IRS, such as IRS reflection design, IRS channel estimation, IRS deployment, etc. For example, the transmit (active) beamforming at the AP and the reflect (passive) beamforming at the IRS need to be jointly designed so that IRS-reflected and non-IRS reflected energy signals  can be coherently combined at the ER to maximize  its received RF power. Besides, channel estimation for IRS associated links is also more practically  difficult as compared to that for the conventional wireless links without IRS, due to the lack of transceiver RF chains at IRS and its large number of elements. Furthermore, although IRS-aided WPT shares similar characteristics with well-studied IRS-aided WIT,   existing designs for WIT  (e.g., \cite{JR:wu2019IRSmaga,JR:wu2018IRS,cui2019secure,guan2019intelligent,chen2019intelligent,
fu2019intelligent,yang2019intelligent,han2018large,huangachievable,yan2019passive,wang2019intelligent,jamali2019intelligent,dongfang2019,JR:wu2019discreteIRS}) may not be applicable to WPT, due to their different design objectives, receiver architectures  and practical  constraints. For example, low-rank  LoS channel  is undesired  for IRS-aided multiple-input-multiple-output (MIMO)/multiuser WIT due to the low spatial multiplexing gain/high co-channel interference, whereas it is  beneficial for IRS-aided WPT since high channel correlation enhances the active/passive  energy beamforming gain of the AP/IRS. Due to the above reasons,  the design of  IRS-aided WIPT systems is also generally different from that of conventional WIPT systems without IRS, thus calling for a new study.

\subsection{Contribution and Organization}
In this paper, we aim to give a tutorial overview on IRS-aided WPT and WIPT systems, by reviewing their state-of-the-art results in the literature as well as providing new ideas to resolve some key challenges in their design and performance optimization, such as  joint active and passive beamforming, communication and energy resource allocation, and IRS channel estimation for low-cost ERs. Moreover,  we  point out important directions worthy of further investigation and other promising topics pertaining to IRS-aided WPT/WIPT for motivating future work. It is worth noting that although there have been a number of  overview/survey articles on IRS-aided WIT (e.g. \cite{JR:wu2019IRSmaga,basar2019wireless,liang2019large,Liaskos2018,elmossallamy2020reconfigurable,huang2019holographic,
wu2021intelligent,di2020smart_JSAC,yuan2020reconfigurable,gong2019towards,bjornson2020reconfigurable}), a tutorial paper dedicated to IRS-aided WPT and WIPT systems is still missing in the literature, to the authors' best knowledge. 

The rest of this paper is organized as follows. Section II addresses   IRS-aided WPT systems and Section III extends the results to IRS-aided WIPT systems, namely IRS-aided SWIPT and WPCN.
In Section IV, we discuss other promising topics related to IRS-aided WPT/WIPT.  Finally, we conclude this paper in Section V.

\subsection{Notation}
In this paper, scalars are denoted by italic letters, vectors and matrices are denoted by bold-face lower-case and upper-case letters, respectively. $\mathbb{C}^{x\times y}$ denotes the space of $x\times y$ complex-valued matrices. For a complex-valued vector $\bm{x}$, $\|\bm{x}\|$ denotes its Euclidean norm,  $[\bm{x}]_n$ denotes its $n$-th element,  $\arg(\bm{x})$ denotes a vector with each element being the phase of the corresponding element in $\bm{x}$, and $\text{diag}(\bm{x})$ denotes a diagonal matrix with the elements in $\bm{x}$ on its main diagonal. The distribution of a circularly symmetric complex Gaussian (CSCG) random vector with mean vector  $\bm{x}$ and covariance matrix ${\bm \Sigma}$ is denoted by  $\mathcal{CN}(\bm{x},{\bm \Sigma})$; and $\sim$ stands for ``distributed as''. For a square matrix $\Ss$, ${\rm{tr}}(\Ss)$ and $\Ss^{-1}$ denote its trace and inverse, respectively, while $\Ss\succeq \bm{0}$ means that $\Ss$ is positive semi-definite, where $\bm{0}$ is a zero matrix of proper size.  For a general matrix $\A$, $\A^*$, $\A^H$,  ${\rm{rank}}(\A)$, and  $\left[{\A}\right]_{i,j}$ denote its conjugate,  conjugate transpose, rank, and the $(i,j)$th entry, respectively. $ \jmath $ denotes the imaginary unit, i.e., $\jmath ^2 = -1 $. $\mathbb{E}(\cdot)$ denotes the statistical expectation. $ \mathrm{Re}\{\cdot\}$ denotes the real part of a complex number. $\otimes$ denotes the Hadamard product.

\section{IRS-aided WPT}
In this section, we consider  IRS-aided WPT systems by addressing their  energy beamforming design and channel estimation issues in practice, followed by discussing other extensions and related works  for further study.

\begin{figure}[!t]
\centering
\includegraphics[width=0.6\textwidth]{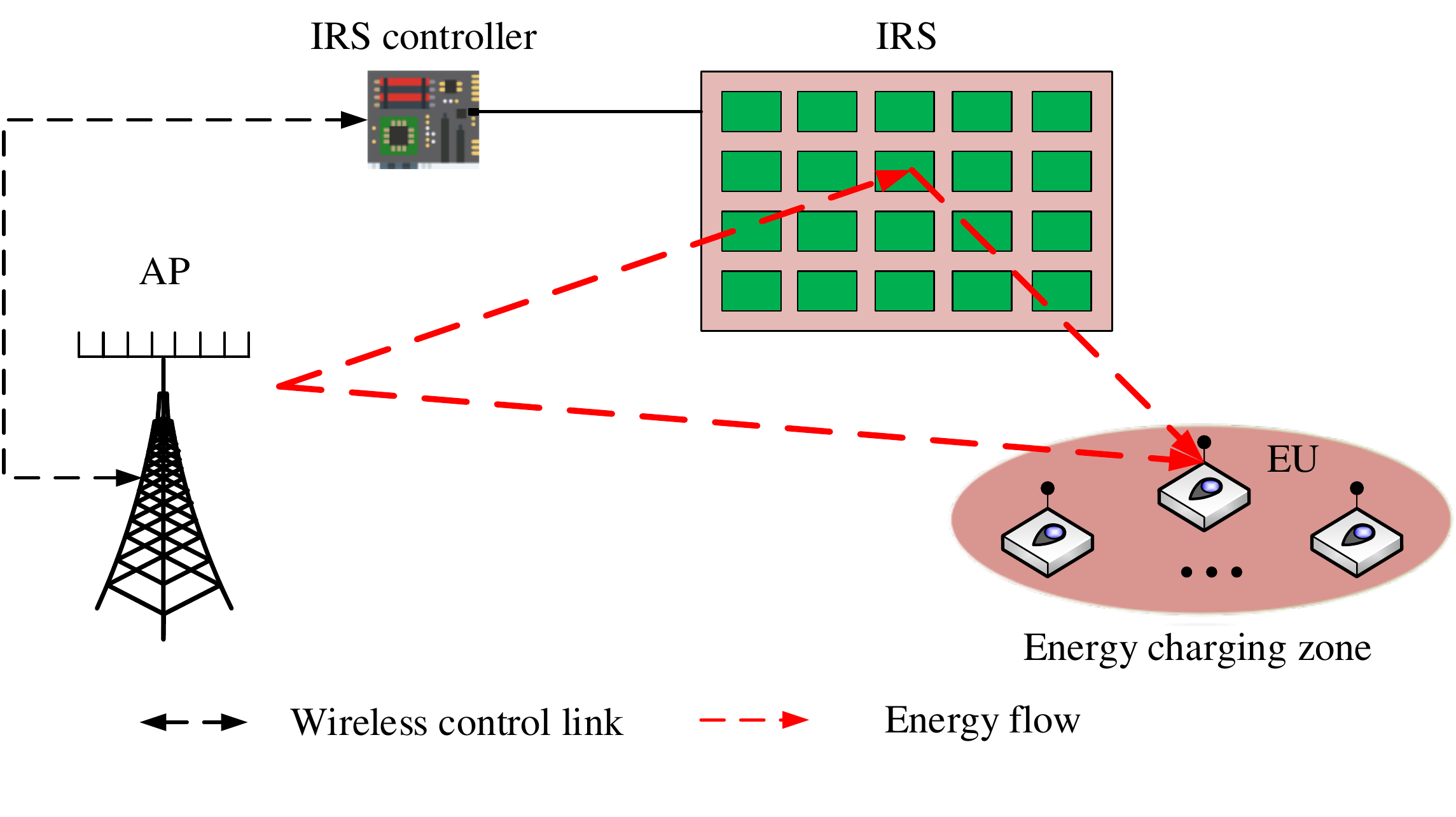}\vspace{-0.05cm} 
\caption{An IRS-aided WPT system. } \label{SII:WPTsystem:model}\vspace{-0.2cm}
\end{figure}

\subsection{Joint Active and Passive Energy Beamforming}
As illustrated in Fig. \ref{SII:WPTsystem:model}, we consider a typical  IRS-aided WPT system consisting of an AP, an IRS, and a set of single-antenna EUs, denoted by the set $\K_{\E}=\{1, \cdots,K_{E}\}$. The AP and IRS are equipped with $M$ antennas and $N$ reflecting elements, respectively, with the set of reflecting elements denoted by $\mathcal{N}$ with $|\mathcal{N}|=N$.
 Consider linear precoding at the AP and assume that each EU is assigned with one dedicated energy beam without loss of generality. Then, the energy signal emitted from the AP can be expressed as
$\mv{x} = \sum_{j\in {\mathcal{K_E}}}{\mv v}_j s_j^{\rm{EH}}$,
where  ${\mv v}_j\in {\mathbb C}^{M\times 1}$ is the precoding vector for EU $j$ and  $s_j^{\rm{EH}}$ is the corresponding energy-carrying signal, satisfying $\mathbb{E}\left(|s_j^{\rm{EH}}|^2\right)=1,\forall j\in \mathcal{K_{E}}$ \cite{xu2014multiuser}.
Denote the total transmit power budget at the  AP by $P$. Then, we have $\mathbb{E}(\mv{x}^H\mv{x}) =  \sum_{j\in {\mathcal{K_E}}}\|{\mv v}_j\|^2 \le P$.
We consider the quasi-static flat fading channel model for all links for the purpose of exposition, and assume that the CSI of all links is known  perfectly to the AP in this subsection to derive the WPT efficiency limit.  The baseband equivalent channels from the AP to  EU $j$,  from  the IRS to  EU $j$, and from the AP to IRS   are denoted by $\bm{g}^H_{d,j} \in \mathbb{C}^{1\times M} $,  $\bm{g}^H_{r,j}\in \mathbb{C}^{1\times N}$, and $\bm{F}\in \mathbb{C}^{N\times M}$, respectively.  Let  $\ttheta  = \text{diag} (\beta_1 e^{\jmath\theta_1}, \cdots, \beta_N e^{\jmath\theta_N})$ denote the (diagonal) reflection-coefficient matrix at the IRS, where   $\beta_n \in [0, 1]$ and $\theta_n\in [0, 2\pi)$ denote the reflection amplitude and  phase shift of the $n$th element, respectively \cite{JR:wu2019IRSmaga,JR:wu2018IRS}.  For simplicity, we set $\beta_n=1$, $\forall n\in \N$ to maximize the signal reflection by the IRS. By ignoring the noise power, the received RF power at EU $j$, denoted by $E_j$, is given by
\begin{align}\label{EH:energy}
E_j=    \sum \limits_{k\in\mathcal{K_E}}|(\bm{g}^H_{r,j}\ttheta \bm{F} +  \bm{g}^H_{d,j}) {\mv v}_k|^2  =   \sum \limits_{k\in\mathcal{K_E}}|(  {\bm u}^H  {\hat \F_j} +  \bm{g}^H_{d,j}) {\mv v}_k|^2
 \triangleq   \sum \limits_{k\in\mathcal{K_E}}|\g^H_j {\mv v}_k|^2,\ \forall j\in \mathcal{K_{E}},
\end{align}
where ${\hat \F_j}=   \text{diag}(\bm{g}^H_{r,j}  )\bm{F}$ and ${\bm u} = [u_1, \cdots, u_N]^H$ with $u_n =e^{\jmath\theta_n}, \forall n$.

To balance the harvested energy amounts at different EUs, we aim to maximize the weighted sum-power received by all the EUs subject to the total transmit power constraint at the AP and the phase-shift constraints at the IRS.  Denote  the energy weight of EU $j$  by  $\alpha_j\geq 0$. A larger value of $\alpha_j$ implies a higher priority for transferring energy to EU $j$ as compared  to other EUs.  Based on  \eqref{EH:energy}, the weighted sum-power received by all the EUs is given by
\begin{align}
\sum \limits_{j\in\mathcal{K_E}} \alpha_j E_j =  \sum\limits_{j\in\mathcal{K_E}}{\mv v}_j^H{\mv S}{\mv v}_j,
\end{align}
where $\Ss =\sum_{j\in  \K_{\E}}\alpha_j\g_j\g^H_j$.  Accordingly,  the joint active and passive energy beamforming optimization problem can be formulated as
\begin{align}
\text{(P1)}: ~~\max_{\{\bm{v}_j\}, \bm{u} } ~~~&\sum_{j\in \K_{\E}} \bm{v}_j^H\Ss\bm{v}_j    \label{eq:obj}\\
\mathrm{s.t.}~~~~& \sum_{j\in \K_{\E}}\|\bm{v}_j\|^2\leq P,  \\
&|u_n|=1, \forall n\in\mathcal{N}.  \label{phase:constraints}
\end{align}
Note that (P1) is a non-convex optimization problem since  the objective function is not jointly concave with respect to $\bm{v}_j$'s and $\bm{u}$. In particular, the same set of IRS phase shifts are shared by all the EUs and thus need to be carefully designed to tradeoff their harvested energy. Nevertheless, it is observed  that by fixing either the phase shifts at the IRS or energy precoders at the AP, the reduced problem of (P1) can be efficiently  solved, which thus motivates  us to employ the alternating optimization (AO) approach to solve (P1) sub-optimally by alternately optimizing $\bm{v}_j$'s and  $\bm{u}$ in an iterative manner until the convergence is achieved.

Specifically, for any fixed $\bm{u}$, it is not difficult to  show that the optimal energy precoder for each EU should be aligned with the principle eigenvector of  $\Ss$  corresponding to its largest eigenvalue \cite{xu2014multiuser}, denoted by ${\bar \vvv}_0(\Ss)$. As such, sending only one common energy beam is optimal for (P1) and the optimal energy precoder is given by
\begin{align}
\vvv^*_j= \vvv_0=\sqrt{P}\bar \vvv_0(\Ss), \forall j.
\end{align}
In particular, if $\alpha_k\gg\alpha_j, \forall j\neq k$, we have $\Ss \approx \alpha_k\g_k\g^H_k$ and   $\vvv^*_0=\sqrt{P}\g_k/\|\g_k\|$, which implies that the AP should steer its energy beam towards EU $k$ so as to maximize the received sum-power. Second, for any fixed $\vvv^*_0$, (P1) is simplified as
\begin{align}\label{optimize:theta2}
\max_{ \bm{u}} ~~~&\sum_{j\in\mathcal{K_E}} \alpha_j | {\bm u}^H {\bm a}_j+ b_j|^2 \\
\mathrm{s.t.}~~~~&|u_n|=1, \forall n\in\mathcal{N}, \label{unit:modulus:constraint}
\end{align}
where $\bm{a}_{j}={\hat \F_j}\bm{v}^*_0$ and  ${b}_{j}=\bm{g}^H_{d,j}\bm{v}^*_0$, $\forall j$.
Although  the unit-modulus constraints in \eqref{unit:modulus:constraint} are non-convex, we observe that the objective function in \eqref{optimize:theta2} is convex with respect to $\uuu$, which thus motivates us to employ the successive convex approximation (SCA) technique to solve it iteratively \cite{wu2019weighted}.  In summary, by  alternately optimizing the energy precoder and phase shifts, the objective value of (P1) is non-decreasing while  it is upper-bounded by a finite value. Furthermore, there is no coupling between the energy precoder and phase shifts in the constraints of (P1). Therefore, based on the results in \cite{hong2015unified}, it can be shown that the above AO-based algorithm is guaranteed to converge to a stationary solution  of (P1).  Note that in the literature, there are also other techniques to optimize IRS phase shifts for problem \eqref{optimize:theta2},  such as  element-wise optimization,  manifold optimization, and semidefinite relaxation (SDR), etc \cite{wu2021intelligent}.

\begin{figure}[!t]
\centering
\includegraphics[width=0.55\textwidth]{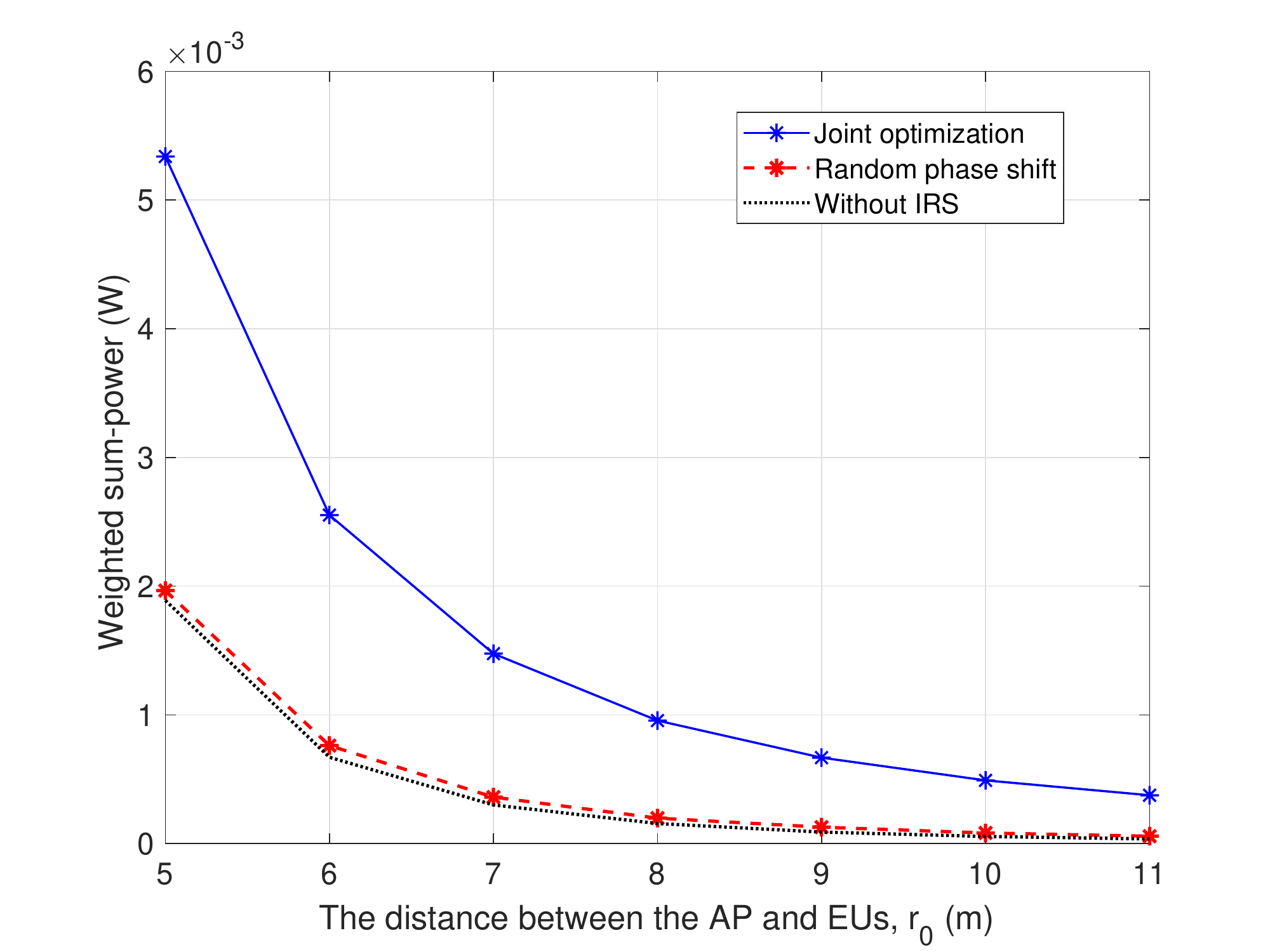}\vspace{-0.1cm}
\caption{Sum-power of EUs versus AP-EU distance. } \label{simulation:distance}\vspace{-0.2cm}
\end{figure}

To show the benefits of IRS brought for WPT, we consider that in Fig. \ref{SII:WPTsystem:model}  all EUs are located between the IRS and the AP with a distance of 2 meters (m) from the IRS. The distance between the AP and EUs is denoted by $r_0$ m. For simplicity, all EUs are assumed to have the same weight, i.e., $\alpha_j=1,\forall j$, and Rayleigh fading is assumed for all channels involved. Besides, the path loss exponents for both AP-IRS and IRS-EU links are set as 2.2 while that of the AP-EU link is set as 3.6.  For comparison, we consider three schemes as follows: 1) Joint active and passive energy beamforming as proposed in the above, 2) Random IRS phase shifts only with optimized transmit beamforming at the AP, and 3) Optimized transmit beamforming at the AP but without IRS.   In Fig. \ref{simulation:distance}, we plot the received  sum-power of EUs versus $r_0$ with $M=4$, $K_E=4$ and $N=100$.  It is observed that by  deploying the IRS around EUs,  their received sum-power is significantly improved as compared to the case without IRS.  In other words, for the same sum-power received at EUs, the WPT range can be  extended without increasing the transmit power at the AP. For example, to achieve the same sum-power of about 0.5 milliwatt (mW), the AP can only cover EUs within a distance of 7 m  in the case without IRS, whereas by deploying the IRS around EUs, the distance  can be improved up to 11 m.  Moreover, the joint passive and active energy beamforming  design significantly  outperforms that  with random IRS phase shifts only, which demonstrates the importance of properly optimized phase shifts at IRS to achieve its energy beamforming gain.

\subsection{Channel Acquisition}
Due to the unique characteristics of IRS and the fundamental differences between WPT and WIT receivers, the channel estimation methods for the traditional WPT systems without IRS or IRS-aided WIT systems may not apply efficiently to the new IRS-aided WPT system. On one hand, compared to traditional WPT systems without IRS, IRS results in a large number of additional channel coefficients with the AP/EUs in the IRS-aided WPT system, which need to be learned with more time and power. On the other hand, different from IRS-aided WIT systems, EUs in IRS-aided WPT systems are usually power-constrained with limited signal processing capability, which thus makes the conventional channel estimation methods relying on receiver pilot transmission and signal processing inefficient or even infeasible. As such, in the following we first give an overview of channel estimation approaches for the conventional WPT systems without IRS and IRS-aided WIT systems, respectively, and then propose new methods for channel acquisition which are customized to  IRS-aided WPT systems.
\subsubsection{Channel Acquisition in WPT Systems Without IRS}
In non-IRS aided WPT systems, CSI could be acquired with similar techniques for WIT systems, but some unique challenges in WPT systems should be considered. Specifically, depending on whether the energy harvesting and communication modules at the EU share the same set of antennas or not, there are two practical EU architectures, namely, shared-antenna architecture and separate-antenna architecture \cite{zengCE}. For the shared-antenna architecture, energy harvesting and communication take place in a time-division manner by using the same antennas with RF switches; while for the separate-antenna architecture, energy harvesting and communication modules use distinct antennas, thus they can be operated concurrently and independently.

For  WPT systems based on the shared-antenna architecture, one straightforward approach to acquire CSI at the AP is by downlink (from AP to EU) training jointly  with uplink (from EU to AP) CSI feedback \cite{WPT_CE}. As different frequency bands can be used for the uplink CSI feedback and downlink training or energy transmission, this scheme applies to both time-division duplex (TDD) and frequency-division duplex (FDD) APs  for separating the uplink/downlink transmissions. However, this scheme requires complex baseband signal processing at EUs, which may not be affordable  for low-cost wireless devices in practice (e.g., RF tags). In addition, the training time increases with the number of antennas at the AP (i.e., $M$). Alternatively, another channel acquisition method is via EU's uplink training by exploiting the channel reciprocity (thus applying to TDD system only), i.e., a fraction of the channel coherence time is assigned to the EU for sending pilot signals to the AP for channel estimation \cite{WPT_CE2}. Its advantage is that channel estimation and feedback are no more required at the EU and the training time is independent of the number of antennas at the AP. However, it requires the EU to transmit pilots, using its limited energy harvested from the downlink. Moreover, a training-energy  tradeoff is involved: too little training results in coarsely estimated channel at the AP and hence a reduced energy beamforming gain, whereas too much training consumes excessive energy harvested by the EU and also leaves less time for energy transmission.

In contrast, for the separate-antenna architecture based WPT systems, the above two channel estimation schemes are both inapplicable due to different communication/energy harvesting  antennas used at the EU, thus the desired CSI for energy harvesting antennas cannot be obtained through estimating the channels with communication antennas. To overcome this issue, a novel channel learning approach that exploits the communication module to feed back the harvested energy levels over time at the energy harvesting module can be adopted \cite{WPT_CE3}. Specifically, the EU measures and encodes the harvested energy levels over different time intervals into bits and feeds them back to the AP. Based on such quantized energy feedback, the AP can adjust the transmit beamforming in subsequent intervals and also obtains refined estimates of the desired channels with energy harvesting antennas. Table \ref{WPT_CE} summarizes the comparison of different channel estimation approaches for non-IRS aided WPT systems.

\begin{table*}[t]
	\scriptsize
	\renewcommand{\arraystretch}{0.9}
	\caption{A Comparison of Channel Estimation Approaches for non-IRS aided WPT Systems.}
	\label{WPT_CE}
	\centering
	\begin{threeparttable}
		\begin{tabular}{|m{1.5cm}<{\centering}|m{3.1cm}<{\centering}|m{2cm}<{\centering}|m{1.4cm}<{\centering}|m{2.1cm}<{\centering}|m{1.3cm}<{\centering}|m{1.6cm}<{\centering}|}
			\hline
			EU architecture & Chanel estimation approach &Uplink/downlink duplexing
			&Sending pilot by EU &Complex signal processing at EU &Feedback from EU &Training time order \\
			\hline
			\multirow{2}{1.5cm} {\centering Shared-antenna }  & Downlink training with uplink feedback \cite{WPT_CE} &TDD, FDD
			& No & Yes & Yes&${\cal O}(M)$\\
			\cline{2-7}   & Uplink training exploiting channel reciprocity  \cite{WPT_CE2}   &TDD     & Yes & No&No&${\cal O}(1)$\\
			\hline
			Separate-antenna  & Channel learning based on EU's energy feedback \cite{WPT_CE3}& TDD, FDD & No &  No & Yes & ${\cal O}(M)$ \\		
			\hline
		\end{tabular}
	\end{threeparttable}
	\vspace{-4mm}	
\end{table*}

\subsubsection{Channel Acquisition in IRS-aided WIT Systems}
Depending on whether the IRS is mounted with sensing devices (receive RF chains) or not, there are two different configurations, termed as semi-passive IRS and fully-passive IRS, respectively, which require  different channel estimation approaches \cite{wu2021intelligent}. With semi-passive IRS,  channel estimation is executed by leveraging  the sensors on IRS to receive pilot signals from the AP/IUs for estimating their respective channels to IRS. Due to close proximity of sensors and IRS reflecting elements, the AP/IU-IRS channels can be constructed approximately  from the estimated channels with sensors by exploiting their strong spatial correlation and applying signal processing techniques, e.g., compressed sensing, data interpolation, and machine learning \cite{IRS_CE1}. Then, based on  the channel reciprocity (which holds for TDD systems only), the CSI of the reverse IRS-AP/IU links can be obtained.

On the other hand, for fully-passive IRS, estimating the AP-IRS and IRS-IU channels separately becomes infeasible. To address this issue, a practically viable approach is to estimate the cascaded AP-IRS-IU channels, based on the training signals from the AP/IU by properly designing the IRS reflection pattern over time \cite{IRS_CE2}. However, the main challenge lies in that, since IRS usually consists of a large number of reflecting elements, the required pilot/training overhead becomes prohibitively high in practice. To reduce channel estimation overhead, an efficient approach is to group adjacent IRS elements into a sub-surface,  referred to as IRS element grouping, whereby only the effective cascaded IU-IRS-AP channel associated with each sub-surface needs to be estimated \cite{IRS_CE3}. Moreover, for IRS-aided multi-user communication, a key observation is that the cascaded IRS channels of all IUs share a common AP-IRS channel, which can be exploited to improve the channel estimation efficiency. For example, one IU can be selected as the reference IU of which the cascaded channel is first estimated. Then, based on this reference IU's CSI, the cascaded channels of the remaining IUs can be efficiently estimated by exploiting the fact that these cascaded channels are scaled versions of the reference IU's cascaded channel, thus only the corresponding low-dimension scaling factors need to be estimated \cite{IRS_CE4}. However, the efficiency of the above reference user based channel estimation is improved most pronouncedly when the number of antennas at the AP (i.e, $M$) is sufficiently large, which may not be the case certain applications  (e.g., WiFi system). For such systems with small $M$, an anchor-assisted channel estimation scheme was proposed in \cite{IRS_CE5} by exploiting dedicated  anchors near the IRS, which is applicable  to TDD systems.  
Moreover, in contrast to the above approaches for directly estimating the cascaded channels, another practical approach is to acquire the channels implicitly via codebook-based beam training, i.e., by simply comparing the received signal power over different IRS reflection patterns from a predefined codebook, each user can find its optimal IRS beam pattern with maximum channel gain and sends its index to the IRS controller  \cite{IRS_CE6}. Note that the channel estimation approaches proposed in \cite{IRS_CE2,IRS_CE3,IRS_CE4,IRS_CE6} apply to both TDD and FDD systems. Specifically, for TDD systems, the estimated uplink CSI can also be used to design IRS reflection for downlink transmissions, while in FDD systems, the uplink CSI and downlink CSI should be estimated independently. A comparison of the above channel acquisition approaches for IRS-aided WIT systems (by using the downlink channel estimation as an illustration) is given in Table \ref{IRS_WIT_CE}.

\begin{table*}[t]
	\scriptsize
	\renewcommand{\arraystretch}{0.9}
	\caption{A Comparison of Channel Acquisition Approaches for IRS-aided WIT Systems (Downlink Case).}
	\label{IRS_WIT_CE}
	\centering
	\begin{threeparttable}
		\begin{tabular}{|m{1.5cm}<{\centering}|m{3.9cm}<{\centering}|m{1.9cm}<{\centering}|m{1.3cm}<{\centering}|m{2cm}<{\centering}|m{1.1cm}<{\centering}|m{1.2cm}<{\centering}|}
			\hline
			IRS configuration & CSI acquisition approach &Uplink/downlink duplexing & Sending pilot by IU & Complex signal processing at IU  & Feedback from IU       &Training time order  \\
			\hline
			Semi-passive IRS  & Using sensors on IRS for channel construction \cite{IRS_CE1}& TDD & Yes &  No & No & ${\cal{O}}(M)$ \\	
			\hline
			\multirow {5}{1.5cm}{\centering Fully-passive IRS}
			&\multirow{2}{4cm} {\centering Cascaded channel estimation by varying IRS reflection pattern  \cite{IRS_CE2} }
			&TDD & Yes &No & No &${\cal{O}}(1)$\\
			\cline{3-7}   &  &FDD & No & Yes & Yes& ${\cal{O}}(M)$\\
			\cline{2-7}
			& \multirow{2}{3cm} {\centering IRS elements grouping based training  \cite{IRS_CE3} }
			&TDD     & Yes & No&No&${\cal{O}}(1)$\\
			\cline{3-7}   &  &FDD & No & Yes & Yes& ${\cal{O}}(M)$\\
			\cline{2-7}
			& \multirow{2}{3cm} {\centering Reference user based channel training \cite{IRS_CE4} }
			&TDD     & Yes & No&No&${\cal{O}}(1/M)$\\
			\cline{3-7}   &  &FDD & No & Yes & Yes& ${\cal{O}}(M)$\\
			\cline{2-7}
			& Anchor-assisted training  \cite{IRS_CE5}   &TDD     & No & Yes&Yes&${\cal{O}}(1)$\\
			\cline{2-7}
			& Codebook-based beam training \cite{IRS_CE6}   &TDD, FDD     & No & No&Yes&${\cal{O}}(1)$\\				
			\hline		
		\end{tabular}
	\end{threeparttable}
	\vspace{-4mm}
\end{table*}

\subsubsection{Efficient Chanel Acquisition for IRS-aided WPT Systems}
\begin{figure}[t]
	\centering
	\vspace{0mm}	
	\subfigure[Training and transmission protocol for single EU setup.]
	{
		\includegraphics[width=6in]{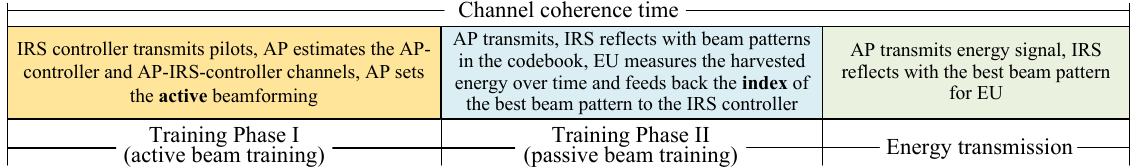}
		\label{fig:CE_singleEU}
	}\vspace{3mm}
	\subfigure[Training and transmission protocol for multi-EU setup.]
	{
		\includegraphics[width=6in]{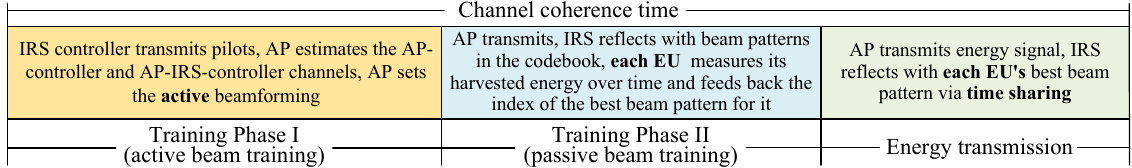}
		\label{fig:CE_multiEU}
	}
	\caption{EU energy measurement based beam training for IRS-aided WPT.}
	\label{CE_WPT}
	\vspace{-5mm}	
\end{figure}

From Table \ref{WPT_CE} and Table \ref{IRS_WIT_CE}, we can draw some useful insights into the efficient channel acquisition design for the new IRS-aided WPT system. Since downlink WPT is of our interest, we consider the CSI acquisition for the downlink in this subsection. First, consider the semi-passive IRS-aided WPT system with separate-antenna EU architecture. In this case, it is infeasible to estimate the IRS-EU channel since distinct antennas are used for communication and energy harvesting at the EU. While for the semi-passive IRS-aided WPT system using shared-antenna EU architecture, the desired AP-IRS-EU channel can be obtained by estimating the two constituting channels with the IRS sensors separately, as in the semi-passive IRS-aided WIT systems \cite{IRS_CE1}. However, note that due to EU's limited transmit power, the IRS should be deployed at the user side to establish a strong short-distance link and thereby reduce the pilot power consumption at the EU. This is in sharp contrast to the case of IRS-aide WIT systems, where deploying the semi-passive IRS at the AP side or at the user side are both applicable \cite{Wu2019TWC}. On the other hand, for the case with a fully-passive IRS, it turns out  that combining the energy measurement feedback from the EU with the codebook-based IRS beam training is a practically efficient method for CSI acquisition in IRS-aided WPT systems, which applies to both EU architectures as well as both TDD and FDD systems.  It also has two further advantages: (i) only low-rate feedback is required without the need of pilot transmission or complex signal processing at the EU; (ii) the IRS-EU channel training time is independent of the number of antennas at the AP. Motivated by the above, in the following we propose new and efficient two-phase beam training schemes for both single-user WPT and multiuser WPT scenarios based on EU's energy measurement feedback in IRS-aided WPT systems, as illustrated in Fig. \ref{CE_WPT}.

{\textbf{Single-user WPT}: }For the purpose of exposition, we consider a fully-passive IRS-aided single EU system based on TDD. As shown in Fig. \ref{fig:CE_singleEU}, the channel coherence time is divided into two training phases (i.e., an active beam training phase and a passive beam training phase) and an energy transmission phase. In Training Phase I, the IRS controller is exploited to send pilots such that the AP can estimate the AP-controller and AP-IRS-controller channels, respectively (for FDD system, these channels can also be obtained at IRS controller based on pilots sent by the AP). Based on such estimated CSI, the active beamforming (i.e., ${\bm v}$) and the passive beamforming (i.e., ${\bm u}^H$) can be optimized, by e.g.,  the AO method proposed in \cite{JR:wu2018IRS}, to maximize the received power at the IRS controller. Specifically, assuming that the AP-controller channel follows arbitrary fading (denoted by ${\bm g}^H_{d,0}$), while the channels between IRS elements and its controller are LoS due to their short distance, which are denoted by $\bm{g}^H_{r,0}=g_0\bm{a}^H({\omega_0^a},{\omega_0^e})$, where $g_0$ denotes the complex-valued path gain of the IRS-controller link, ${\omega_0^a} \in [0, \pi]$ and ${\omega_0^e} \in [0, \pi]$ denote the azimuth AoD and the elevation AoD from the IRS to the controller, respectively, and $\bm{a}^H({\omega_0^a},{\omega_0^e})$ represents the array response vector of IRS. As such, the optimal $\bm v$ and ${\bm u}^H$ can be obtained as
\begin{align}
(\bm v_0,{\bm u}_0^H)=\arg \max_{ {\bm v}, {\bm u}^H} |{\bm g}^H_{d,0}{\mv v}+{{\bm u}^H}\text{diag}(\bm{g}_{r,0}^H ) \bm{F}  {\mv v}|^2.
\label{bf_single}
\end{align}
As shown in \cite{JR:wu2018IRS}, $(\bm v_0,{\bm u}_0^H)$ must satisfy that $\angle ({\bm g}^H_{d,0}{\mv v}_0)=\angle ({{\bm u}_0^H}\text{diag}(\bm{g}_{r,0}^H ) \bm{F}  {\mv v}_0)$. By denoting $\phi_0=\angle ({\bm g}^H_{d,0}{\mv v}_0)$, $\bm {h}_0=\bm{F}  {\mv v}_0$ and $\bm {\varphi }_0=\angle (\bm{g}_{r,0}^H \odot \bm{h}_0^T)$, we have $\bm u_0^H=e^{j({\phi_0-\bm {\varphi }_0})}$.

Note that the active beamforming $\bm v_0$ is kept unchanged in the following passive beam training and energy transmission phases, since it has been aligned with the AP-IRS channel; while the remaining task is to find a proper passive beam pattern aligning with the IRS-EU and BS-EU channels. Specifically, in Training Phase II, the received power at the EU is given by
\begin{align}
E_u=|{\bm g}^H_{d,u}{\mv v}_0+{{\bm u}^H} (\bm{g}_{r,u}^*  \odot \bm{h}_0)|^2,
\end{align}
where ${\bm g}^H_{d,u}$ is the AP-EU channel, and $\bm{g}^H_{r,u}=g_u\bm{a}^H({\omega_u^a},{\omega_u^e})$ is the IRS-EU LoS channel (assuming a short distance between them). Moreover, $g_u$ denotes the complex-valued path gain of the IRS-EU link, while ${\omega_u^a} \in [0, \pi]$ and ${\omega_u^e} \in [0, \pi]$ denote the azimuth AoD and the elevation AoD from the IRS to the EU, respectively. Let $\phi_u=\angle ({\bm g}^H_{d,u}{\mv v}_0)$ and $\bm {\varphi }_u=\angle (\bm{g}_{r,u}^H  \odot \bm{h}_0^T)$. It can be easily observed that the optimal passive beamforming vector for EU is given by $\bm u^H_u=e^{j(\phi_u-\bm {\varphi }_u)}$. Denote the phase difference by $\Delta\phi=\phi_u-\phi_0$ and $\Delta\bm {\varphi }=\bm {\varphi }_u-\bm {\varphi }_0$. Then $\bm u^H_u$ can be expressed as ${\bm u^H_u}={\bm u^H_0}\odot e^{j(\Delta\phi+\Delta\bm {\varphi })}$, i.e., $\bm u^H_u$ can be obtained from $\bm u^H_0$ with a common phase rotation $\Delta\phi$ (to align the IRS reflected channel with the direct channel) and an element-wise phase rotation $\Delta\bm {\varphi }$ (to adjust the passive beam direction). Furthermore, since $\Delta\bm {\varphi }$ is determined by $\Delta\bm {\varphi }=\angle (\bm{g}_{r,u}^H  \odot \bm{h}_0^T)-\angle (\bm{g}_{r,0}^H  \odot \bm{h}_0^T)=\angle(g_u)-\angle(g_0)+\angle(\bm{a}^H({\omega_u^a},{\omega_u^e}))-\angle(\bm{a}^H({\omega_0^a},{\omega_0^e}))$, ${\bm u^H_u}$ can be rewritten as ${\bm u^H_u}={\bm u^H_0}\odot e^{j(\Delta\bar\phi+\Delta\bar{\bm {\varphi }})}$, where $\Delta\bar\phi=\angle ({\bm g}^H_{d,u}{\mv v}_0)-\angle ({\bm g}^H_{d,0}{\mv v}_0)+\angle(g_u)-\angle(g_0)$ is a random phase difference due to the fading channel, while $\Delta\bar{\bm {\varphi }}=\angle(\bm{a}^H({\omega_u^a},{\omega_u^e}))-\angle(\bm{a}^H({\omega_0^a},{\omega_0^e}))$ is a deterministic phase difference vector only dependent on the locations of IRS (controller) and EU.

Motivated by the above result, the passive beam training codebook at IRS can be constructed as ${\bf W}=\{e^{j\phi_i}{\bf w}_l^H,i=1,...,N_\phi,l=1,...,N_\varphi\}$, where $N_B=N_\phi N_\varphi$ is the number of beam patterns in total, while $\phi_i=\frac{2\pi}{N_\phi}(i-1)$ and ${\bf w}_l^H\in {\mathbb C}^{1\times N}$ are designed for searching the common phase rotation and the element-wise phase rotation, respectively. Assuming that the IRS is composed of $N = N_x \times N_z$ reflecting elements placed in the $x-z$ plane, we can rewrite ${\bf w}_l^H$ as ${\bf w}_l^H=[1,e^{j\varphi_x^{l_1}},...,e^{jN_x\varphi_x^{l_1}}] \otimes [1,e^{j\varphi_z^{l_2}},...,e^{jN_z\varphi_z^{l_2}}]$, i.e., the Hadamard product of the horizontal and vertical IRS beam training vectors, where  $\varphi_x^{l_1}=\frac{2\pi}{N_1}({l_1}-1)$, $\varphi_x^{l_2}=\frac{2\pi}{N_2}({l_2}-1)$, $l_1=1,...,N_1$,  $l_2=1,...,N_2$ and $N_\varphi=N_1 \times N_2$. Note that each horizontal/vertical beam pattern has a main-lobe with beam width $2\pi/N_1$ and $2\pi/N_2$, respectively, for achieving the corresponding maximum beam gain of $N_1$ or $N_2$. With the above designed codebook $\bf W$, the AP consecutively sends multiple training symbols while the IRS changes its reflecting pattern in $\bf W$ over different training symbols. Meanwhile, the EU measures its harvested energy amount in each symbol duration and feeds back the index of the beam pattern achieving the largest harvested energy amount to IRS controller at the end of training. Then, the best beam pattern for the EU can be set at the IRS to facilitate WPT in the remaining energy transmission phase. Note that the above protocol can be modified to work for semi-passive IRS by replacing the role of IRS controller with that of sensors in Training Phase I.

{\textbf{Multiuser WPT}: The above proposed beam training scheme for a single EU can be extended to the general multi-EU setup shown in Fig. 1. As shown in Fig. \ref{fig:CE_multiEU}, Training Phase I is the same as that in the single EU case, while in Training Phase II, each EU measures its harvested energy amount over time and feeds back the index of its best beam pattern to the IRS controller. At last, in the WPT phase, the IRS reflects with each EU's best beam pattern via proper time sharing among EUs to cater to their different weights in the weighted sum-energy maximization. Moreover, the proposed beam training scheme can also be extended to the WPT system with multiple EU clusters, each of which is served by an IRS nearby. Specifically, in the active beam training phase, multiple IRS controllers transmit orthogonal pilots to the AP simultaneously such that the AP can estimate all the AP-controller channels and AP-IRS-controller channels; then, the AP transmits one or more energy beams towards these controllers based on the estimated CSI. By fixing the active beamforming in the passive beam training phase, each IRS searches its best beam pattern by using the method in the above. 

As shown in Fig. \ref{IRS_WPT_CE} (with the same system setup as Fig. 2), the proposed beam training based scheme achieves more pronounced active/passive joint beamforming gains with the increasing number of beam patterns in the codebook (i.e.,  $\bf W$). Note that in the proposed scheme, the active beamforming is designed based on maximizing the received power at the IRS controller and then fixed for the subsequent passive beam training, thus incurring certain performance loss as compared to their joint optimization based on AO with the perfect CSI.

\begin{figure}[t]
	\centering
	\vspace{0mm}		
	\includegraphics[width=3.4in]{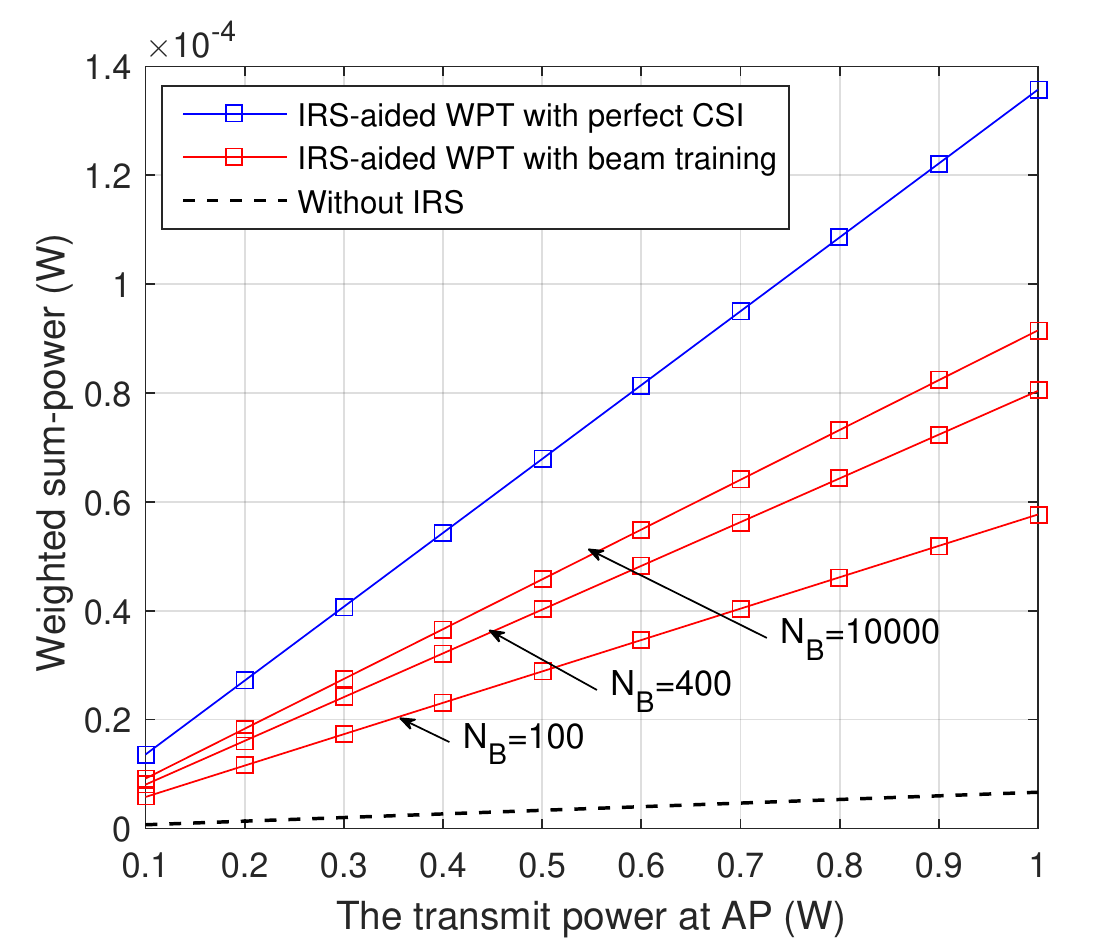}
	\caption{Harvested sum-power versus transmit power at the AP, where  the number of EUs is $K_E=4$ and that of beam patterns in the codebook is $N_B=$100, 400, and 1000, respectively.}
	\label{IRS_WPT_CE}
	\vspace{-5mm}	
\end{figure}

At last, since locations of the AP, IRS and IRS controller are fixed in practice, the channel coherence time of the direct/reflect channels of the IRS controller with AP  is much larger than that of the user-associated channels. This implies that the active beam training phase only needs to be conducted over long periods. Furthermore, the passive beam training time can also be reduced if the location of EU is known. Specifically, in this case, the element-wise phase rotation vector $\Delta\bar{\bm {\varphi }}=\angle(\bm{a}^H({\omega_u^a},{\omega_u^e}))-\angle(\bm{a}^H({\omega_0^a},{\omega_0^e}))$ can be directly obtained without training, since the array response vectors of IRS depend only on the azimuth AoD and elevation AoD from the IRS to the controller/EU, which can be obtained given their location information. As a result, only the common phase rotation needs to be trained, which significantly reduces the passive beam training overhead.

\subsection{Extensions and Future Work}
Despite the promising benefit of IRS for WPT systems, there have been only  limited works in this area  \cite{zhao2021WPT,wu2019jointSWIPT,haiyuan2020IRS,mishra2019channel}, which considered different design objectives for the IRS-aided WPT system in Fig. \ref{SII:WPTsystem:model}. 
In contrast to (P1), maximizing the minimum harvested energy among all EUs was considered in \cite{zhao2021WPT} where active and passive beamformers are jointly optimized by leveraging asymptotic results, which, however, are suboptimal in general for the case with finite $M$ and/or $N$.  In \cite{wu2019jointSWIPT}, the transmit power at the AP was minimized by jointly optimizing active and passive beamforming subject to a given set of quality-of-service (QoS) requirements on  the individual harvested energy amounts of EUs. 
 An interesting finding in \cite{wu2019jointSWIPT}  is that the deployment of IRS in WPT systems not only effectively reduces the transmit power (or equivalently improve WPT efficiency/coverage) but also simplifies the precoding design at the AP by reducing the number of active energy beams needed, despite that multiple energy beams are generally needed to meet EUs' QoS constraints. To reduce the system implementation complexity, a joint design considering the low-cost constant-envelope analog beamforming at the AP was investigated in \cite{haiyuan2020IRS} for IRS-aided WPT systems. Nevertheless, research on this new paradigm is still in an early stage and there are still many important issues that are open and worthy of further investigation. In the following, we discuss some promising topics to motivate future work.

In the preceding subsections, we adopt the RF power received at EUs (i.e.,  the input of the energy harvester at EUs) as the performance limit for WPT, while a more practically relevant metric is actually the output DC power of each energy harvester. In the literature, there are generally three different energy harvester models, namely the diode linear model,  diode non-linear model, and  saturation nonlinear model \cite{clerckx2018fundamentals}. While our proposed designs are applicable to IRS-aided WPT systems with the most commonly used diode linear model with a constant energy conversion efficiency at the EU, they can be different for the two more realistic non-linear models. In particular, besides the received signal power, the output DC power of an RF energy harvester under the diode non-linear model also depends on the waveform of received signals, e.g., deterministic multisine waveform and modulated orthogonal frequency division multiplexing (OFDM) waveform \cite{clerckx2016waveform}. As such, a joint signal waveform and IRS passive beamforming design was studied in \cite{feng2021waveform} for both IRS-aided single and multiuser WPT systems where the waveform complex weights and IRS phase shifts are alternately optimized based on EUs' individual frequency-selective CSI.  Note that the continuous IRS phase shift was assumed in \cite{feng2021waveform}, whereas it remains unclear about the effects of other practical IRS reflection models with e.g. amplitude-dependent/discrete IRS phase shifts   \cite{abeywickrama2020intelligent} \cite{wu2018IRS_discrete} on the waveform/passive beamforming design and WPT performance. Moreover, the non-linear model effects on the active/passive beam training design are also worth studying in future work.

Channel acquisition for IRS-aided WPT in frequency-selective channels is generally more challenging than in their  frequency-flat counterpart since more channel coefficients are resulted for both AP-EU direct channels and AP-IRS-EU reflected channels due to the multi-path delay spread and the resultant convolution of time-domain impulse responses of AP-IRS and IRS-EU multi-path channels in the cascaded AP-IRS-EU channel. Moreover, note that although the channels are frequency-selective, the IRS reflection coefficients are time-selective only but frequency-flat, which thus cannot be flexibly designed for different frequencies (e.g., different sub-carriers in OFDM systems). Fortunately, the number of OFDM sub-carriers is typically much larger than the maximum number of delayed paths in practical wireless systems, thus there exists great redundancy that can be exploited for designing OFDM-based pilot symbols to efficiently estimate the channels of multiple users at the same time \cite{IRS_CE7}. It is very interesting to observe that  the previously discussed approaches for high-efficiency multiuser narrow-band channel estimation, such as IRS element grouping \cite{IRS_CE3} and BS-IRS common channel exploitation \cite{IRS_CE4}\cite{IRS_CE5}, etc., rely on excessive spatial channels for training overhead reduction, while the method proposed in \cite{IRS_CE7} exploited the redundancy of OFDM sub-carriers to achieve this goal. Motivated by this, it is also possible to fully exploit the redundancy in both antennas and OFDM sub-carriers to further improve the training efficiency of frequency-selective channels. However, the above works considered the IRS-aided WIT only, while efficient channel acquisition for IRS-aided broadband WPT systems still needs further investigation, by exploiting the IRS controller and EU's energy measurement feedback as in our proposed beam training method for narrow-band channels.

Last but not the least,  the previous discussion assumes that the IRS is deployed at a given location. Generally speaking,  different IRS deployment strategies may lead to drastically different realizations/distributions of the end-to-end channels and thus significantly affect the system performance. 
Despite that for the IRS-aided WPT system in Fig. \ref{SII:WPTsystem:model} with $M = 1$ and $K_E = 1$ shares the same optimal deployment strategy as the IRS-aided WIT system (i.e., placing the IRS close to the AP or EU \cite{wu2021intelligent}), it may not hold when the channel acquisition issue or the MIMO/multiuser case is considered, due to the fundamental differences in design objective and receiver architecture  between WPT and WIT. For example, deploying an IRS with LoS rather than rich-scattering links with both the AP and EUs is practically preferred for IRS-aided WPT system since it not only leads to smaller path loss and hence lower signal attenuation over distance, but also results in strong correlation among the channels among EUs and thus enhanced passive beamforming gain.
This is in sharp contrast to the IRS deployment strategy in IRS-aided MIMO/multiuser WIT systems, where the IRS deployment is more involved as it needs to balance the LoS and non-LoS components in its reflected channels to achieve both high spatial multiplexing and beamforming gains.
Moreover, considering the scenario where massive number of EUs are randomly distributed in future wireless networks, a general hybrid deployment scheme by properly placing IRSs near both APs and EUs is expected to provide more flexibility and superior performance for WPT over deploying them near the AP or EUs only, especially when the cooperative beamforming among distant IRSs is explored \cite{zheng2021double}. This thus gives rise to various new challenges that need to be tackled, such as IRS placement, IRS-EU/AP association, multi-hop IRS beamforming, etc.

\section{IRS-aided WIPT}
In this section, we study IRS-aided SWIPT and WPCN, respectively, by extending the results in Section II.

\begin{figure}[!t]
\centering
\includegraphics[width=0.6\textwidth]{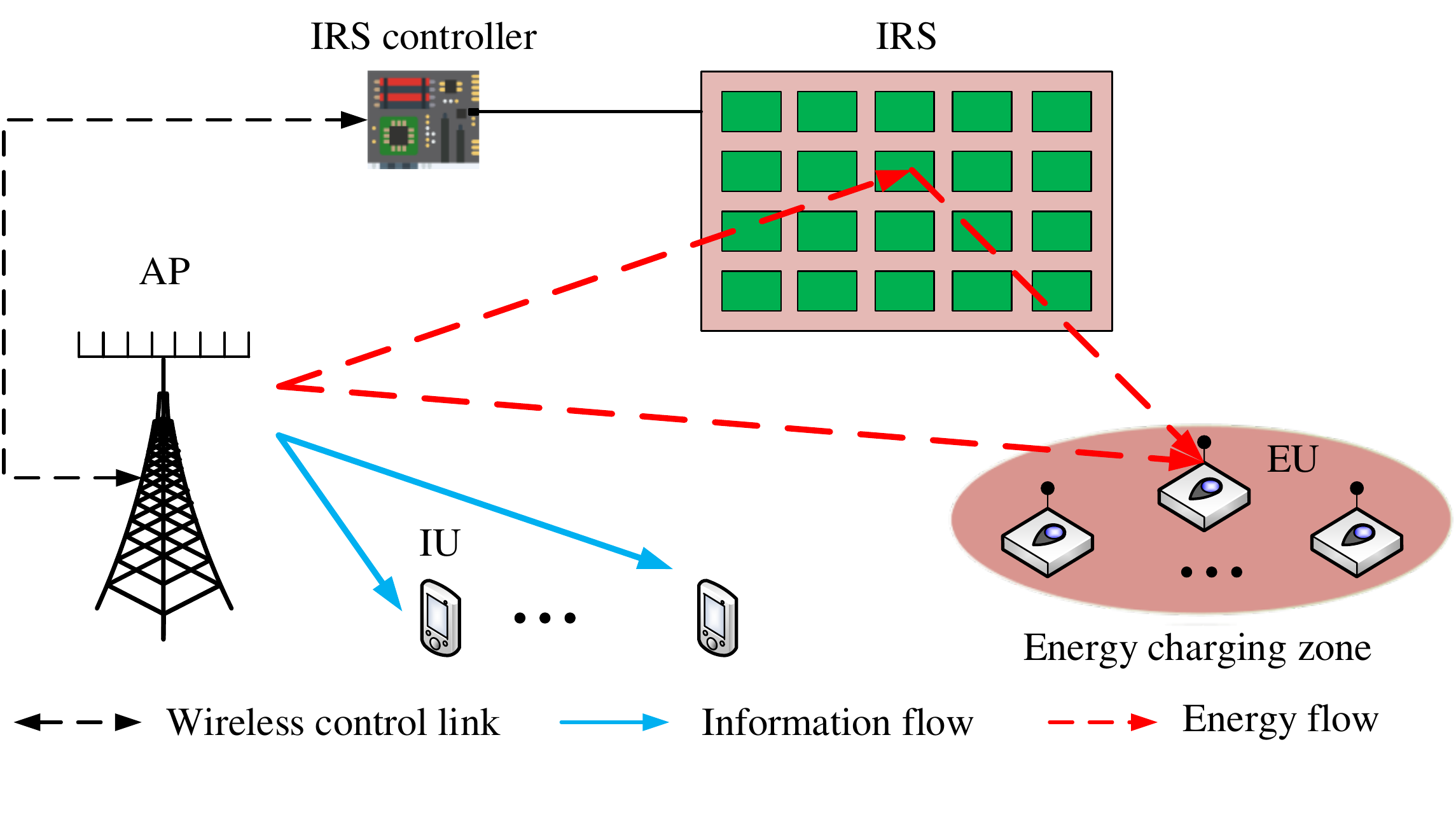} 
\caption{An IRS-aided SWIPT system. } \label{SIII:SWIPT:system:model}
\end{figure}

\subsection{IRS-aided SWIPT}
Recall that SWIPT aims to achieve concurrent information and energy transmission efficiently over a given spectrum. Thus, there exists a non-trivial tradeoff in allocating communication resources such as power, time and bandwidth to achieve a balanced performance between information transmission rates and harvested energy amounts at the IUs and EUs, respectively. In this subsection, we first present the joint energy and information beamforming design for SWIPT and then address the channel acquisition issues, followed by the extensions as well as other promising topics for future work.

\subsubsection{Joint Energy and Information Beamforming}
As shown in Fig. \ref{SIII:SWIPT:system:model}, we consider a typical IRS-aided SWIPT system where a set of single-antenna IUs, denoted by $\K_{\II}=\{1, \cdots,K_{I}\}$, are added into the WPT system in Fig. \ref{SII:WPTsystem:model}.  Similar to Section II,  linear precoding  is employed at the AP and each IU/EU is assigned with one dedicated information/energy beam. As such, the  signal  transmitted from the AP is expressed as
\begin{align}
\mv{x} = \sum_{i\in {\mathcal{K_I}}}{\mv w}_i s_i^{\rm{ID}} +\sum_{j\in {\mathcal{K_E}}}{\mv v}_j s_j^{\rm{EH}}, \label{equa:jnl:1}
\end{align}
where ${\mv w}_i\in {\mathbb C}^{M\times 1}$ is the precoding vector for IU $i$ and its corresponding information-bearing signal is denoted by $s_i^{\rm{ID}}$, with $s_i^{\rm{ID}} \sim \mathcal{CN}(0,1), \forall i\in \mathcal{K_{I}}$.
Accordingly, the total transmit power constraint at the AP becomes $\mathbb{E}(\mv{x}^H\mv{x}) = \sum_{i\in {\mathcal{K_I}}}\|{\mv w}_i \|^2 +\sum_{j\in {\mathcal{K_E}}}\|{\mv v}_j\|^2 \le P$.
Other system assumptions are the same as those in Section II-A and thus omitted here for brevity.

The baseband equivalent channels from the AP to IU $i$ and from the IRS to IU $i$ are denoted by  $\bm{h}^H_{d,i}\in \mathbb{C}^{1\times M}$ and  $\bm{h}^H_{r,i}\in \mathbb{C}^{1\times N}$, respectively.     Then, the signal received at IU $i$ can be written as
\begin{align}
{y}_i^{\rm{ID}} = (\bm{h}^H_{r,i}\ttheta\bm{F}+\bm{h}^H_{d,i}) \mv{x} + z_i  =  ({\bm u}^H  {\check \F_i} + \bm{h}^H_{d,i}) \mv{x} + z_i  \triangleq  {\mv h}^H_i\mv{x} + z_i,\ \forall i\in \mathcal{K_{I}}, \label{equa:jnl:2}
\end{align}
where  $ {\check \F_i}  = \text{diag}(\bm{h}^H_{r,i}  )\bm{F}$ and $z_i\sim \mathcal{CN}(0,\sigma_i^2)$ is the additive white Gaussian noise (AWGN). Under the assumption that IUs do not possess the capability of cancelling the interference caused by energy signals,  the signal-to-interference-plus-noise ratio (SINR) at IU $i$, $i\in \mathcal{K_{I}}$, is given by
\begin{align}\label{eq:SINR}
\text{SINR}_i = \frac{|{\bm{h}}^H_i\bm{w}_i |^2}{\sum\limits_{ k\neq i, k\in \K_{\II} }|{\bm{h}}^H_i\bm{w}_k |^2 + \sum\limits_{ j \in \K_{\E} }|{\bm{h}}^H_i\bm{v}_j |^2   + \sigma^2_i}.
\end{align}
On the other hand, by ignoring the noise power, the received RF power at EU $j$, denoted by $E_j$, is given by
\begin{align}\label{SIII:EH:energy}
E_j= \sum\limits_{k\in\mathcal{K_I}}|\g^H_j {\mv w}_k|^2  + \sum \limits_{k\in\mathcal{K_E}}|\g^H_j {\mv v}_k|^2,\ \forall j\in \mathcal{K_{E}},
\end{align}
where $\g^H_j = {\bm u}^H  {\hat \F_j} +  \bm{g}^H_{d,j}$ as in Section II.


We aim to maximize the weighted sum-power received by EUs subject to the transmit power constraint at the AP,  unit-modulus phase-shift constraints at the IRS, and  individual SINR constraints at different IUs, given by $\gamma_i, i\in \K_{\II}$.  Based on  \eqref{SIII:EH:energy}, the
weighted sum-power received by all the EUs is given by
\begin{align}
\sum \limits_{j\in\mathcal{K_E}} \alpha_j E_j =  \sum\limits_{i\in\mathcal{K_I}}{\mv w}_i^H{\mv S}{\mv w}_i  + \sum\limits_{j\in\mathcal{K_E}}{\mv v}_j^H{\mv S}{\mv v}_j,
\end{align}
with  $\Ss =\sum_{j\in  \K_{\E}}\alpha_j\g_j\g^H_j$ defined in Section II.  Accordingly,  the joint energy and information beamforming optimization problem is formulated as
\begin{align}
\!\!\!\!\text{(P2)}: \max_{\{\bm{w}_i\}, \{\bm{v}_j\},\bm{u}} & \sum_{i\in \K_{\II}} \bm{w}_i^H\Ss\bm{w}_i + \sum_{j\in \K_{\E}} \bm{v}_j^H\Ss\bm{v}_j   \label{eq:obj}\\
\mathrm{s.t.}~~~~&{\text{SINR}}_{i}\geq \gamma_i, \forall i \in \K_{\II}, \label{P1:SINRconstrn}\\
&\sum_{i\in \K_{\II}}\|\bm{w}_i\|^2 + \sum_{j\in \K_{\E}}\|\bm{v}_j\|^2\leq P,  \\
&|u_n|=1, \forall n\in\mathcal{N}.  \label{phase:constraints}
\end{align}

Note that (P2) is more challenging to solve as compared to (P1) since the IRS phase shifts are further coupled with both energy and information precoders in the SINR constraints in \eqref{P1:SINRconstrn} besides in the harvested power in the objective function.  Furthermore, since  information signals for IUs can also be exploited for energy harvesting at EUs  as shown in \eqref{SIII:EH:energy}, it remains unknown whether dedicated energy beams  are required to maximize the  weighted sum-power of the IRS-aided SWIPT system. Note that this question was \emph{partially} answered for the conventional SWIPT system without IRS in  \cite{xu2014multiuser} by showing that $\bm{v}_j={\bm 0}, \forall j\in \mathcal{\K_\E}$, under the assumption that the channels of all users (both EUs and IUs) are statistically independent/uncorrelated, whereas it still remains open for the case with arbitrary user channels such as correlated/LoS channels. Furthermore, such statistically independent/uncorrelated  channel assumption may not hold  for EUs in the new IRS-aided SWIPT system, as IRS usually results in correlation among the AP-IRS-user effective channels of different EUs via tuning its phase shifts to maximize the EUs' harvested power as shown in Section II. Fortunately, by exploiting the structure of the optimization problem, \cite{wu2019weighted} rigorously proved that even under arbitrary user channels, sending information beams is sufficient for achieving  the optimality of (P1). This result not only resolves the open problem in \cite{xu2014multiuser} completely, but also provides new insights into the optimal beamforming design in SWIPT systems with IRS, which  greatly simplifies the AP precoding design, especially for the case with large $K_E$.  The intuitive explanation of this result is that sending dedicated energy beams not only consumes transmit power at the AP but also causes interference to IUs, and thus should be avoided. By applying this result, (P2) is reduced to
\begin{align}
\max_{\{\bm{w}_i\}, \bm{u}} & \sum_{i\in \K_{\II}} \bm{w}_i^H\Ss\bm{w}_i    \label{eq:obj}\\
\mathrm{s.t.}~~~~&  \frac{|{\bm{h}}^H_i\bm{w}_i |^2}{\sum\limits_{ k\neq i, k\in \K_{\II} }|{\bm{h}}^H_i\bm{w}_k |^2+ \sigma^2_i}
\geq \gamma_i, \forall i \in \K_{\II}, \label{P1:SINRconstrn}\\
&\sum_{i\in \K_{\II}}\|\bm{w}_i\|^2  \leq P,  \\
&|u_n|=1, \forall n\in\mathcal{N}. \label{phase:constraints}
\end{align}
Note that this problem is similar to the non-convex joint active and passive beamforming design problem for IRS-aided WIT systems in \cite{JR:wu2018IRS,wu2021intelligent}, except a different objective function. Existing works  have developed a number of approaches to tackle such problems,  including SDR based AO,  penalty-based method, deep learning based method, alternating direction method of multipliers (ADMM), etc \cite{wu2019weighted,wu2019jointSWIPT,yu2020optimal,feng2020deep,sun2020towards}.

 \begin{figure}[!t] 
\centering
\includegraphics[width=0.5\textwidth]{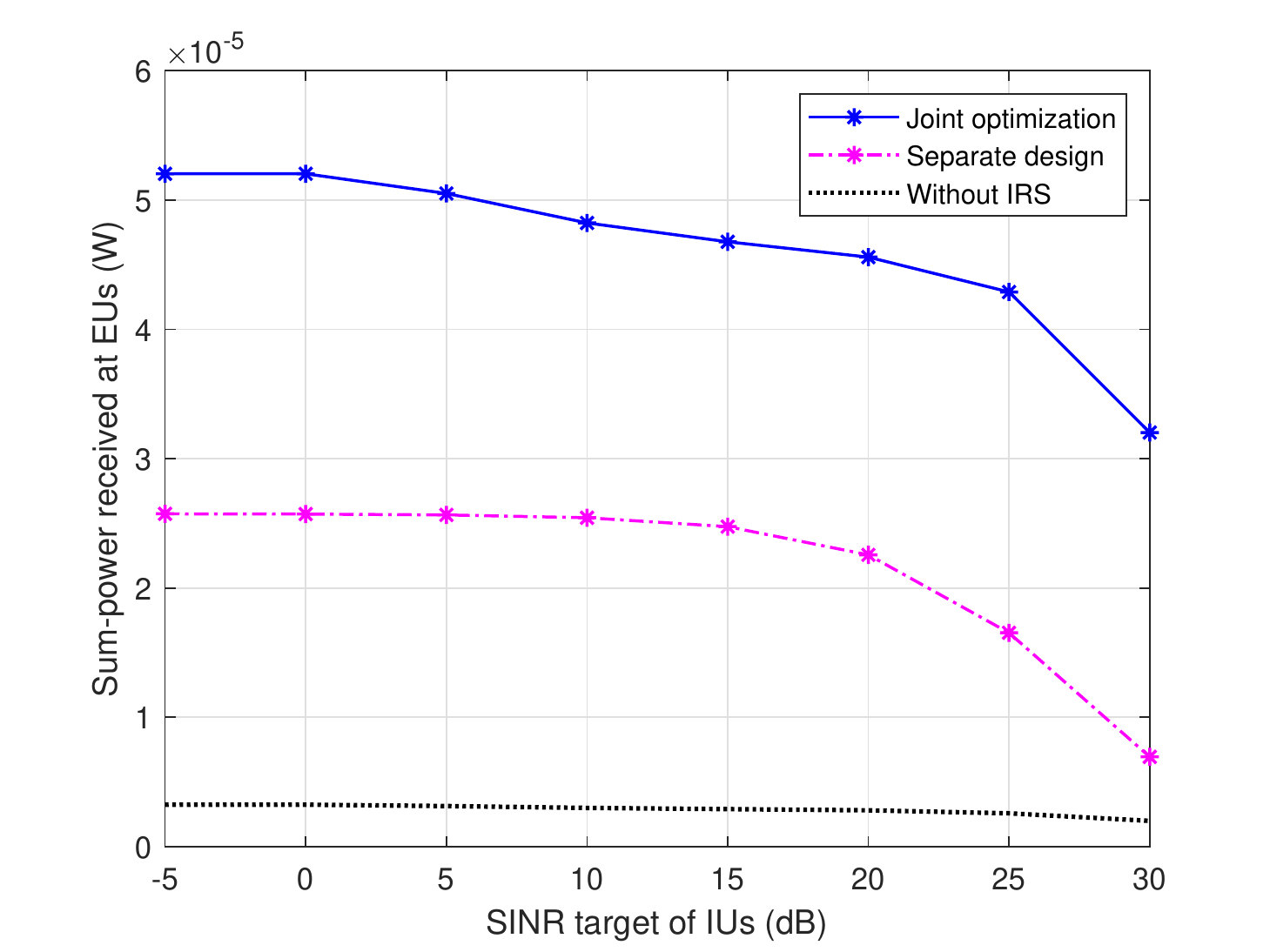}\vspace{-0.1cm}
\caption{Information-energy tradeoff region for SWIPT systems with or without IRS. } \label{simulation:tradeoff}\vspace{-0.2cm}
\end{figure}

As an example, we consider the same setup as in Section II-A by adding IUs which are 50 meters away from the AP.   In  Fig. \ref{simulation:tradeoff}, we plot the tradeoff region between the received  sum-power of EUs and the achievable common SINR of IUs with $M=8$, $K_I=4$, $K_E=4$, and $P=40$ dBm.  Besides the case without using  IRS, the scheme in \cite{xu2014multiuser} with separately designed information/energy beams is also adopted for comparison. For this scheme, the information beams are first designed to minimize the total transmit power required for satisfying all the SINR constraints of IUs, while the remaining AP transmit power is used to send one  energy beam, denoted by ${\bm v}_0$, to maximize the received  sum-power of EUs, subject to the constraint without causing any  interference to all IUs; while the IRS passive beamforming design is same for both cases with IRS. From Fig. \ref{simulation:tradeoff}, it is observed that the achievable power-SINR region of the SWIPT system can be significantly enlarged by deploying the IRS. For example, without compromising the SINR of  IUs, the  received RF power at EUs is greatly improved. 
Furthermore, one can observe that the separate design with one dedicated energy beam suffers considerable performance loss as compared to the joint beamforming design (i.e., with information beams at the AP only as in problem (20)).

\subsubsection{Channel Acquisition}
Next, we extend the energy measurement based beam training scheme proposed for IRS-aided WPT systems in Section II, which requires only energy measurements and low-rate feedbacks of EUs, to the more complex system of IRS-aided SWIPT. Considering the setup in Fig. \ref{SIII:SWIPT:system:model} as an example, we propose a new channel acquisition and information/energy transmission protocol shown in Fig. \ref{CE_SWIPT}, by employing the separate AP information/energy beam design approach for ease of practical implementation. Specifically, each channel coherence time is divided into three phases, which are elaborated as follows. In the first active beam training phase, we exploit the IUs and IRS-controller to transmit pilots for AP, which estimates the AP-IU/controller and AP-IRS-controller channels. Then, based on the estimated CSI, the active beamforming at the AP is optimized, which includes information beams one for each IU and an energy beam to the IRS controller. Next, in the passive beam training phase, the AP sends information signals to IUs and in the meanwhile pilots using the designed energy beam to IRS while the IRS consecutively reflects with different passive beam patterns in a predefined codebook. By comparing the harvested energy amounts over time, each EU can select the best beam pattern and then feeds back its index to the IRS controller. At last,  the AP transmits information and energy signals simultaneously with the fixed active beamforming, while the IRS varies the passive beam patterns over time by properly time sharing  EU's individual best beam patterns.
\begin{figure}[t]
	\centering
	\vspace{0mm}	
	\includegraphics[width=6.5in]{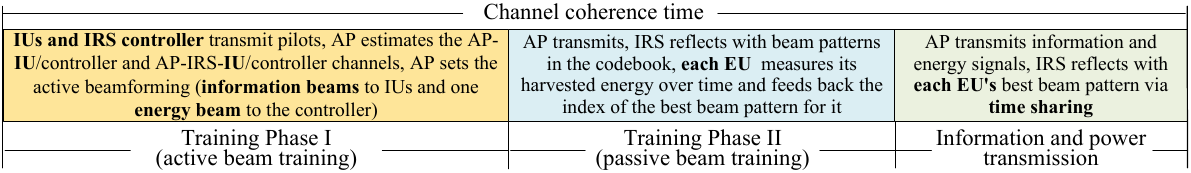}
	\caption{Energy feedback based passive beam training for IRS-aided SWIPT.}
	\label{CE_SWIPT}
	\vspace{-2mm}	
\end{figure}


\subsubsection{Extensions and Future Work}
Research on IRS-aided SWIPT systems has received increasing attention recently due to their appealing benefit for improving the rate-energy performance of IUs and EUs by fundamentally unlocking the energy transmission efficiency limit for EUs. In addition to the works discussed above, we present some other related works and point out in the following promising topics worth investigating in future work.

Besides separate IUs and EUs shown in Fig. \ref{SIII:SWIPT:system:model}, IoT devices may possess a co-located receiver in practical IRS-aided SWIPT systems, i.e., with both information decoding and energy harvesting circuits at the same node. In the literature, time switching (TS) and power splitting (PS) are two commonly adopted designs to implement them with different  rate-energy tradeoffs. Recently,  a few works have studied IRS-aided SWIPT systems with PS-based receivers \cite{zargari2021max,li2020joint,zargari2020energy,zhao2020intelligent} with different objectives such as system energy efficiency maximization, DC power maximization, and transmit power minimization  at the AP. In particular, \cite{zhao2020intelligent} extended the work in  \cite{feng2021waveform} for WPT to a SWIPT system by jointly optimizing transmit power waveform as well as the active and passive beamforming.  However, it considered the single user scenario while more research is needed to investigate the more general multiuser scenario. Although SWIPT systems with PS-based receivers have been studied, their  counterparts with TS-based receivers are also practical and require a further study. Furthermore, since the optimal IRS deployment strategies for WPT and WIT are generally different as discussed in Section II-C, it is worth investigating how to place IRSs in SWIPT systems with different types of receivers, such as IUs and EUs coexisting in the same cluster or EUs with different TS and PS receivers. The joint design of channel acquisition, active/passive beamforming as well as power/information waveform is also worth pursuing.




%


Information security is another critical challenge in conventional SWIPT systems without IRS. Specifically, EUs which are closer to the AP generally have much better channels than IUs and can easily eavesdrop the information sent to IUs. Existing physical layer techniques to address this issue include transmit beamforming with artificial noise (AN)/jamming at the AP \cite{guan2019intelligent}, whereas they may become inefficient in the challenging scenario when  EUs lie in similar direction of IUs or are located near them in practice, if the IUs are not enabled to cancel the interference caused by the energy/AN signals. Fortunately, this issue can be efficiently tackled by deploying IRSs in the vicinity of IUs/EUs and properly designing IRS passive reflections to increase/reduce the achievable rate of IUs/EUs, thus significantly enlarging the achievable secrecy rate-energy region \cite{hehao2020intelligent}. Although we have shown that energy beams are not needed in (P2) for the IRS-aided SWIPT system with perfect CSI, the result may be different for the problem subject to the new secrecy rate constraints at IUs, energy constraints at EUs, or with imperfect CSI. This is because energy beams can be leveraged by IRSs as an effective AN to deliberately interfere with EUs to prevent them from  eavesdropping the IUs' messages, especially when IRSs are deployed near the EUs.  Moreover, given a total number of reflecting elements, it remains unknown how to optimally assign them among different IRSs placed near the AP, IUs, and EUs to achieve the optimal secrecy rate-energy region.

\subsection{IRS-aided WPCN}
WPCN enables low-power wireless devices to both harvest energy and  transmit information from/to the same AP. Different from SWIPT where energy and information flow in the same direction (downlink), in WPCN,  energy and information flow in opposite  directions (i.e., downlink and uplink, respectively). Furthermore, the uplink information transmission rate critically depends o the harvested energy amount in the downlink, which share the same time and frequency resources. In this subsection, we first present the joint passive beamforming and resource allocation design for IRS-aided WPCN and then discuss how to solve its CSI acquisition issue. Last, we present other works/extensions pertinent to this topic.

\begin{figure}[!t]
\centering
\includegraphics[width=3.6in]{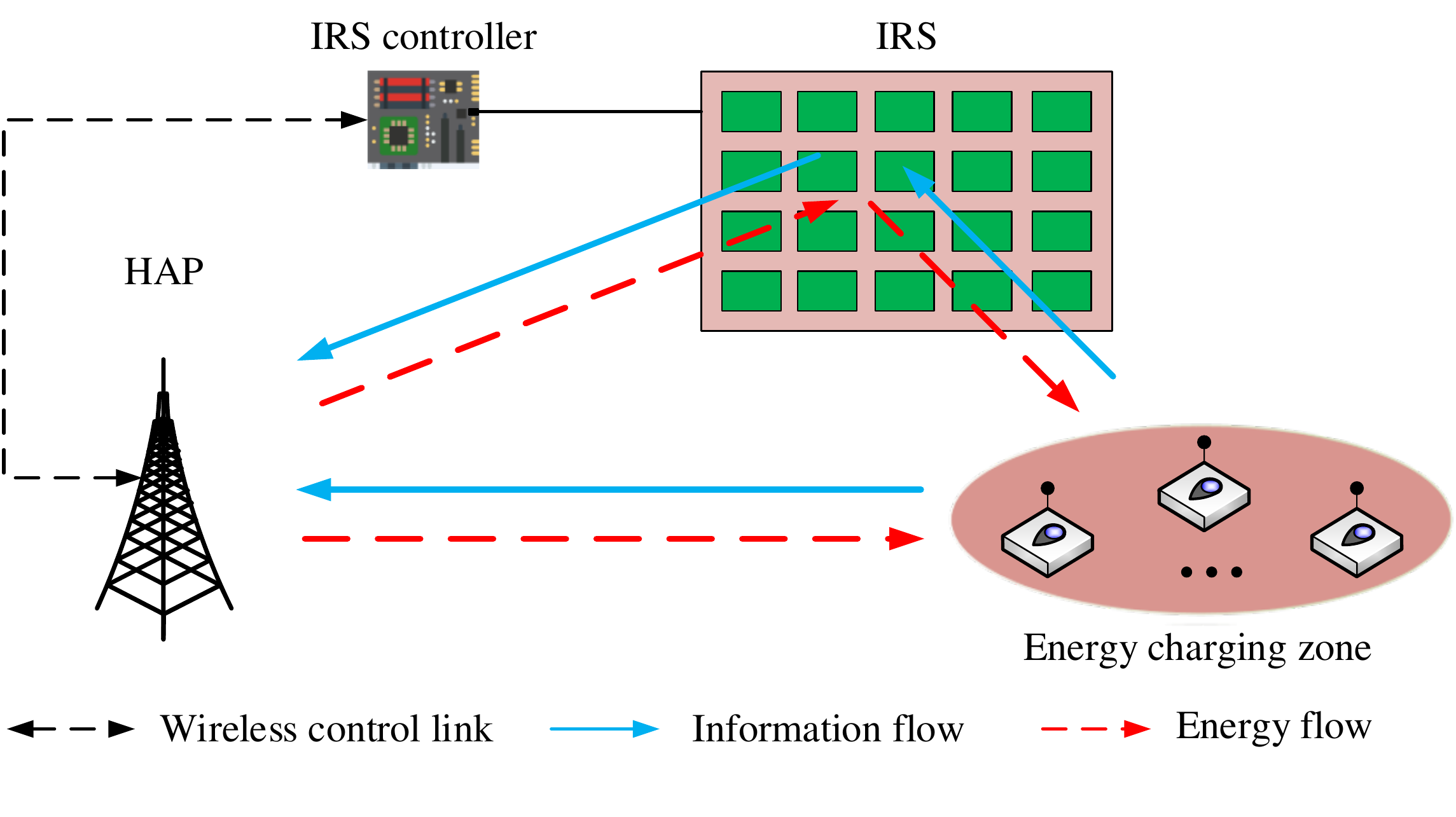}
\caption{An IRS-assisted WPCN with downlink WPT and uplink WIT.}\label{system:model:WPCN}
\end{figure}

\subsubsection{Joint Passive Beamforming and Resource Allocation}

As illustrated in Fig. \ref{system:model:WPCN}, we consider a typical IRS-aided WPCN, where an IRS with $N$ reflecting elements is deployed to assist the downlink WET as well as uplink WIT between a single-antenna  AP and  $K$ wireless-powered devices. 
The AP and all devices are assumed to operate over the same frequency band and  the total available transmission time is denoted by $T_{\max}$.  Similar to Section II, the quasi-static flat-fading channel model is adopted such that all the channel coefficients remain constant during $T_{\max}$.  The well-known ``harvest then transmit'' protocol \cite{ju14_throughput} is adopted, i.e., in each channel coherence interval, wireless powered devices first harvest energy from the downlink signal sent by the AP and then use the harvested energy to transmit information signals to the AP in the uplink.    
 The equivalent baseband  channels  from the AP to IRS, from the IRS to device $k$, and from the AP to device $k$ are denoted by $\bm{g}\in \mathbb{C}^{N\times 1}$, $\bm{h}^H_{r,k}\in \mathbb{C}^{1\times N}$, and ${h}^H_{d,k}\in \mathbb{C}$, respectively, where $k = 1, \cdots,K$.

{During downlink WPT,} the AP transmits with a constant transmit power denoted by $P$  to broadcast energy signals for a duration denoted by $\tau_0$, with  $\tau_0<T_{\max}$.
By considering the diode linear energy harvesting model in Section II-C, the amount of  energy harvested at device $k$ via downlink  WPT  can be written as
\begin{align}\label{eq3}
E^h_k&=\eta_kP|h^H_{d,k} +  \bm{h}^H_{r,k}\ttheta_0 \bm{g}|^2\tau_0   \nonumber  \\
&=\eta_kP|h^H_{d,k} +   \bm{q}_k^H \uuu_0|^2\tau_0,
\end{align}
where $\eta_k \in (0,1]$ denotes the RF-to-DC power conversion efficiency of device $k$, $\q_k= \bm{h}^H_{r,k}   \text{diag}(\bm{g})$ and $\uuu_0 = [e^{\jmath\theta^0_1},..., e^{\jmath\theta^0_N}]^T$ denotes the IRS's phase-shift vector  in downlink WPT. During uplink WIT, each wireless powered device independently transmits its own  information signal to the AP with transmit power $p_k$ for a duration of $\tau_k$. Without loss of generality, the IRS can reconfigure its phase-shift vectors in uplink WIT, each for one of the $K$ devices. Accordingly, the achievable information rate of device $k$ in bits per second per Hertz (bps/Hz) is given by
\begin{align}\label{eq6}
r_k=\tau_k \log_2\left(1+\frac{p_k |h^H_{d,k} + \q^H_k \vvv_k|^2}{\sigma^2}\right),
\end{align}
where  $\uuu_k = [ e^{\jmath\theta^k_1}, \cdots,  e^{\jmath\theta^k_N}]^T$  denotes the  IRS's phase-shift vector for device $k$ during its uplink WIT, and  $\sigma^2$ is the AWGN power at the AP.

 Our objective is to maximize the {weighted sum rate} of all devices  by jointly optimizing the time allocations for downlink WPT and uplink WIT, transmit powers of devices, and IRS phase shifts for both WPT and WIT phases. Accordingly, the joint passive beamforming and resource allocation optimization problem is formulated as
\begin{subequations} \label{probm10}
\begin{align}\label{eq10}
\text{(P3)}: ~~ \mathop {\mathrm{max} }\limits_{{\tau_{0},\{\tau_{k}\},\{p_{k}\}, \uuu_0,\{\uuu_{k}\} } }~~ &\sum_{k=1}^{K} \alpha_k\tau_k  \log_2\left(1+\frac{p_k |h^H_{d,k} + \q^H_k \uuu_k|^2}{\sigma^2}\right) \\
\mathrm{s.t.} ~~~~~~~&  {p_k}\tau_k\leq    \eta_kP|h^H_{d,k} +   \bm{q}_k^H \uuu_0|^2\tau_0, ~ \forall\, k, \label{P1:EH} \\
& |[\uuu_0]_n|=1,  n=1,\cdots, N, \label{P1:eq:modulus1} \\
& |[\uuu_k]_n|=1,  n=1,\cdots, N, \forall k, \label{P1:eq:modulus2}\\
& \tau_{0}+\sum_{k=1}^{K}\tau_k\leq T_{\mathop{\max}},  \label{SecII:eq402} \\
& \tau_{0}\geq0, ~  \tau_k\geq  0,  ~p_k\geq  0, ~\forall k.  \label{SecII:eq403}
\end{align}
\end{subequations}
where  $\alpha_k$ denotes the non-negative rate weight of device $k$ accounting for its priority.  First, it can be shown that the optimal phase shifts during device $k$'s WIT should align IRS-reflected signal with that over the AP-user link to maximize the received signal power at the AP, which are given by  $[\uuu^*_k]_n= e^{\jmath( \arg\{h^H_{d,k}\} - \arg\{ [\q^H_k]_n\}  )}, \forall n$. Let $ | h^H_{d,k} +  \q^H_k \uuu_0  | = |   {\bar \q}^H_k \bar \uuu_0   | $, where ${\bar\uuu}_0 = [ \uuu^H_0 \: 1]^H$ and $ {\bar \q}^H_k  =  [  {\q}^H_k  \: h^H_{d,k}] $, $\Q_k={\bar \q}_k{\bar \q}^H_k$, and $\bm{V}_0=\bm{\bar{u}}_0\bm{\bar{u}}_0^H$ which needs to satisfy  $\bm{V}_0\succeq \bm{0}$ and ${\rm{rank}}(\bm{V}_0)=1$. Then, for constraints \eqref{P1:EH}, it follows that $| {\bar \q}^H_k \bar \uuu_0   |^2 =  {\bar \q}^H_k \bar \uuu_0 \bar \uuu^H_0 {\bar \q}_k  = {\rm{Tr}}(\Q_k\V_0  ), \forall k$.  Furthermore,   constraints \eqref{P1:eq:modulus1}, i.e., $|[\uuu_0]_n|=1$,  are equivalent to  $[\V_0]_{n,n} = 1, n=1,\cdots,N+1$.
For (P3), we apply a change of variables as  $e_k=\tau_kp_k$ and $\W_0= \tau_0\V_0$, which yields
\begin{subequations} \label{probm10}
\begin{align}\label{SecIII:P1}
\mathop {\mathrm{max} }\limits_{{\tau_{0},\{\tau_{k}\},\{e_{k}\}, \W_0 } }~~ &\sum_{k=1}^{K}  \alpha_k \tau_k\log_2\left(1+ \frac{e_k}{\tau_k}\frac{\gamma_k}{\sigma^2}   \right) \\
\mathrm{s.t.} ~~~~~~~&  e_k \leq \eta_kP  {\rm{Tr}}(\Q_k\W_0  ), ~ \forall k,  \\
& [\W_0]_{n,n} = \tau_0, ~ n=1,\cdots, N+1, \label{P6:C9}\\
& {\rm{rank}}(\W_0)=1,  \label{SecIII:rank1}\\
& \eqref{SecII:eq402},  \tau_{0}\geq0, ~  \tau_k\geq  0,  ~e_k\geq  0, ~\forall k,
\end{align}
\end{subequations}
where $\gamma_k \triangleq  |h^H_{d,k} + \q^H_k \uuu^*_k|^2$. With the rank-one constraint \eqref{SecIII:rank1} relaxed, problem \eqref{probm10} becomes a convex semidefinite program (SDP) and can be  solved optimally by using standard convex optimization solvers such as CVX, which also provides a performance upper bound for evaluating other suboptimal solutions to (P3). To obtain a feasible rank-one solution to (P3), Gaussian randomization can be applied to the solution obtained as detailed in \cite{wu2018IRS,JR:wu2018IRS}. Another approach to solve problem  \eqref{probm10} is to  transform constraint \eqref{SecIII:rank1} into an equivalent one based on the largest singular value of $\W_0$ and then solve the resulting problem iteratively by applying the penalty method \cite{wu2019jointSWIPT}.



\begin{figure}[!t]
\centering
\subfigure[Performance comparison.]  {\includegraphics[width=3.2in, height=2.5in]{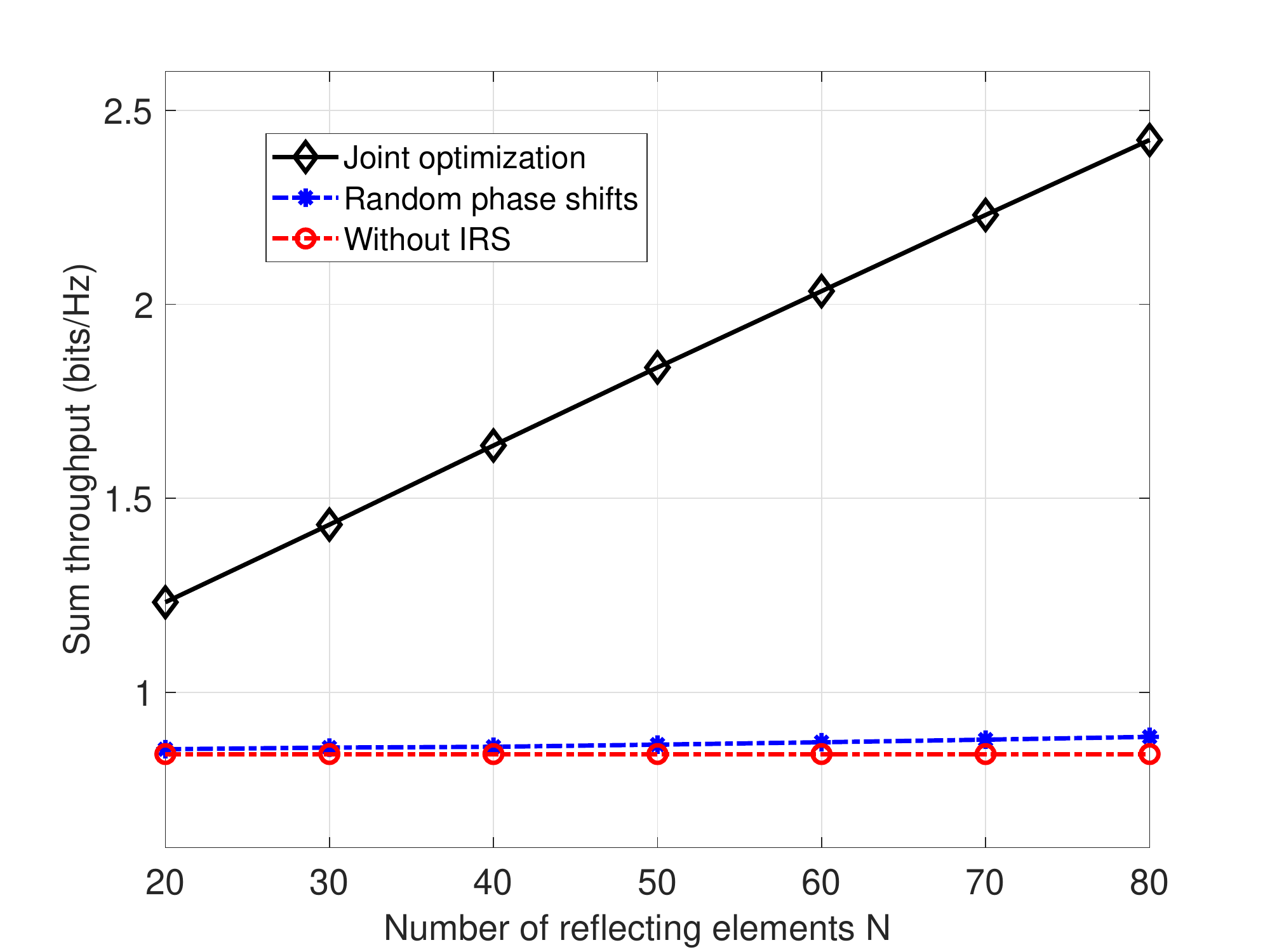}\label{N:versus:rate}}
\subfigure[Impact of $N$ on downlink WPT duration.]{\includegraphics[width=3.2in, height=2.5in]{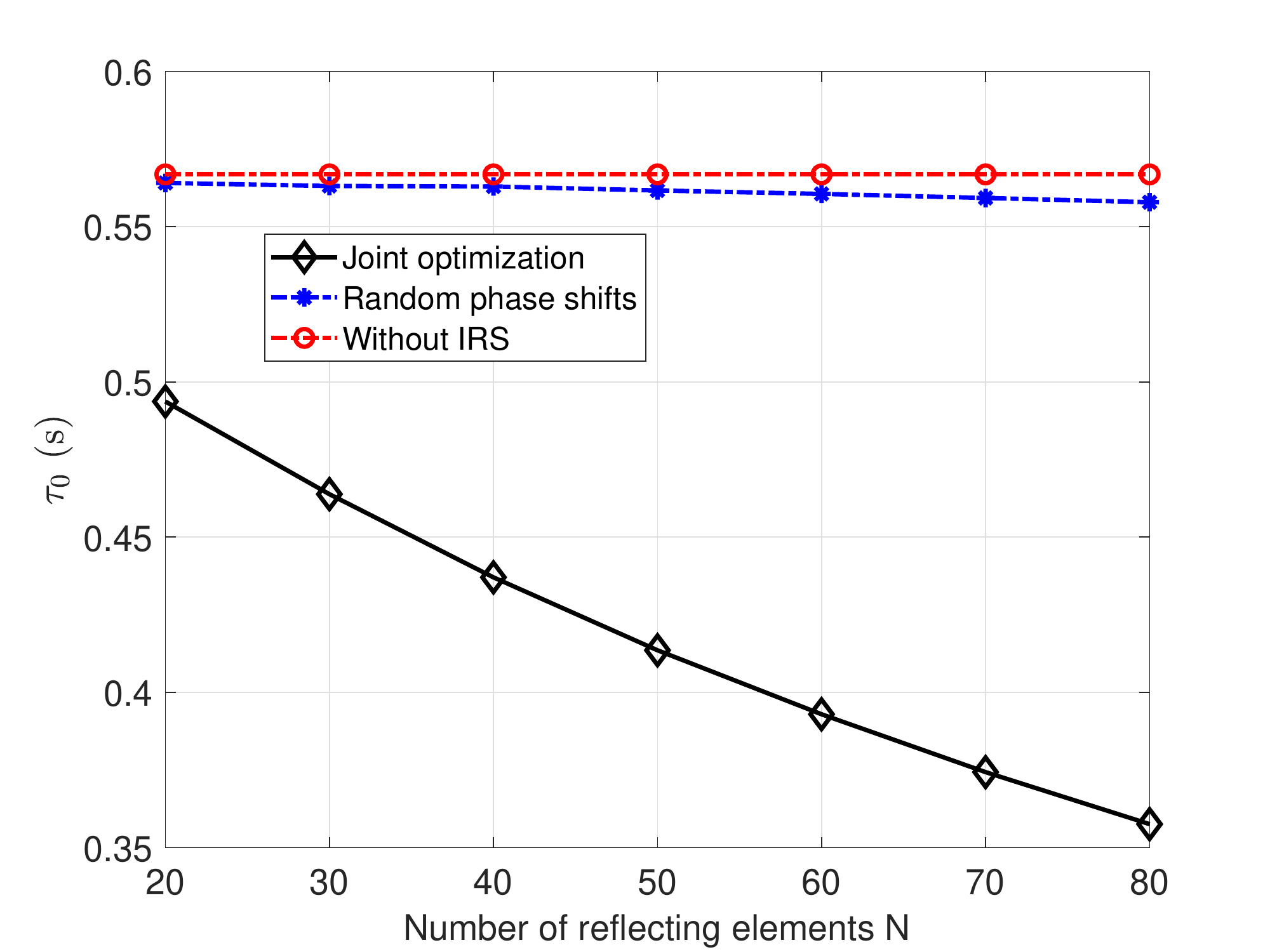}\label{N:versus:tau0}}
\caption{Simulation results for WPCN with or without IRS. } \label{pb}
\end{figure}

Based on the same simulation setup in Section II,  we plot in Fig. \ref{N:versus:rate} the sum throughput versus the number of IRS elements with $K=10$, $P=40$ dBm,  $\sigma^2=-85$ dBm, $\alpha_k=1,\forall k$. and $\eta_k=0.8,\forall k$. We consider the following schemes for comparison: 1) Joint beamforming and time allocation optimization via solving problem \eqref{probm10} and applying Gaussian randomization,  2) Random IRS phase shifts with optimized time allocation, and 3) Optimized time allocation but without IRS. It is first observed from Fig. \ref{N:versus:rate} that the system sum throughput is significantly improved by deploying an IRS with the joint  IRS beamforming and resource allocation as compared with the scheme without IRS, whereas the sum throughput with random IRS phase-shift only achieves marginal gains over that without IRS. This thus demonstrates the effectiveness of IRS for WPCN as well as the importance of joint beamforming and resource allocation  design. It is worth pointing out that besides throughput improvement in Fig. \ref{N:versus:rate},  the total energy consumption at the AP also decreases, which is denoted by $E_{\rm AP}=P\tau_0$. To show this explicitly, we plot  the optimized downlink WPT duration $\tau_0$  versus $N$ in Fig.  \ref{N:versus:tau0}.  One can observe that  the optimized $\tau_0$ in the IRS-aided WPCN decreases as $N$ increases with the joint beamforming and time allocation design, which implies reduced energy consumption at the AP.
This also leaves devices more time for uplink WIT, which helps increase the WIT sum-throughput. As such, exploiting IRS for WPCN brings double benefits as it not only enhances the system throughput but also lowers the energy consumption at the AP.


\begin{figure}[t]
	\centering
	\vspace{0mm}	
	
	\includegraphics[width=6.5in]{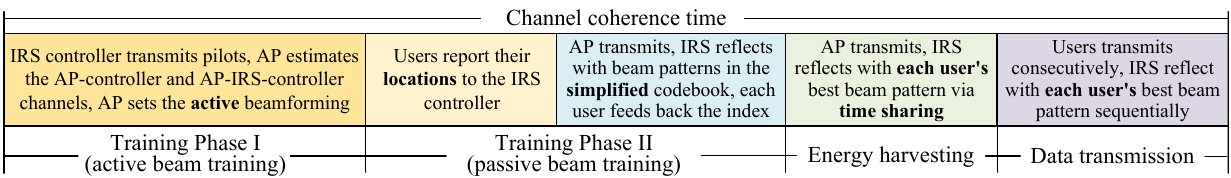}
	\caption{Energy feedback based passive beam training for IRS-aided WPCN.}
	\label{CE_WPCN}
	\vspace{-2mm}	
\end{figure}

\subsubsection{Channel Acquisition}
Considering an IRS-aided WPCN shown in Fig. \ref{system:model:WPCN}, the energy feedback based beam training scheme proposed for the IRS-aided WPT in Section II can be extended to IRS-aided WPCN. Specifically, as shown in Fig. \ref{CE_WPCN}, each channel coherence time is divided into four phases, namely the active beam training phase, passive beam training phase, energy harvesting phase, and data transmission phase. The first three phases, i.e., active/passive beam training and downlink WET are the same as those the IRS-aided WPT  and thus omitted for brevity. The last phase, i.e., uplink information transmission, is added, during which the users send their respective information via time-division multiple access (TDMA) while the IRS sets accordingly its phase shift vector as the optimal one for each user. Note that based on the received information rates of different users, the AP can adjust the time allocations for WPT and WIT to achieve its design objective (e.g., weighted sum-rate maximization).


\subsubsection{Extensions and Future Work}
In addition to the above discussions, there are other important topics/works on IRS-aided WPCN, which are briefly discussed as follows, to motivate future work.

It has been assumed that WPCN employs TDMA for uplink WIT, while other practical multiple access schemes such as space-division multiple access (SDMA) and non-orthogonal multiple access
(NOMA) \cite{wu2018spectral,gong2019optimal} may be considered as well, which generally require different active/passive beamforming and resource allocation designs. For example,  a similar IRS-aided WPCN as Fig. \ref{system:model:WPCN} employing NOMA was studied in \cite{wu2021irs} with the objective of maximizing the sum throughput via joint time allocation and IRS phase shifts optimization. Interestingly, it unveiled that the same IRS phase-shift vector should be adopted for both downlink WPT and uplink WIT in the optimal solution, which is different from the case of TDMA/SDMA based WPCN \cite{zheng2020joint,YuanZheng2020irs,gong2021throughput,li2021robust}.  Moreover, the effects of different multiple access schemes on IRS channel acquisition methods and deployment strategies are important problems to pursue in future work, e.g, how to deploy IRSs to balance the LoS and NLoS links in IRS-aided WPCN employing SDMA/NOMA.

\section{Other Related Topics}

\subsection{IRS-aided WPT/WIPT Meets Autonomous Vehicles}
Mobile charger, e.g., that mounted on autonomous ground vehicle (AGV) or  unmanned aerial vehicle (UAV), is a promising solution to extend the WPT coverage and improve the WPT efficiency by exploiting its mobility control to shorten the charger-user distance. However, new challenges arise when serving a large number of users distributed in a wide area. In such cases, the charger needs to move to be sufficiently close to each user for efficient WPT, which results in not only excessive power consumption but also complex path planning problem to solve. To overcome this issue, deploying IRSs in the network can be an effective solution. For example, by deploying an IRS near a cluster of users, the mobile charger may not need to move too close to these users as the IRS helps enhance the WPT link strength for them, as shown in Fig. \ref{future}(a). On the other hand, instead of employing IRSs at fixed locations in WPT/SWIPT/WPCN systems, which only have fixed half-space reflection coverage and may not have LoS links with users at all locations, mounting IRS on AGV/UAV as mobile IRS can significantly enhance its coverage in terms of both angle and distance by adjusting its facing and/or location based on the locations of IUs/EUs aided by it. 


Mobile charger/AP assisted by fixed IRSs and mobile IRS-assisted charger/AP at fixed locations give rise to many interesting problems to solve. For example, the channel acquisition issue becomes more complicated because of the rapidly time-varying channels due to the mobile AP/IRS. As such, more efficient channel acquisition methods are needed to tackle this issue. Moreover, the placement or trajectory design for mobile APs/IRSs should be jointly optimized with the active/passive beamforming and resource allocation to maximize the system performance while minimizing the AGV/UAV energy consumption to make best use of their limited onboard battery.

\begin{figure}[t]
	\centering
	\vspace{0mm}	
	
	\includegraphics[width=6.5in]{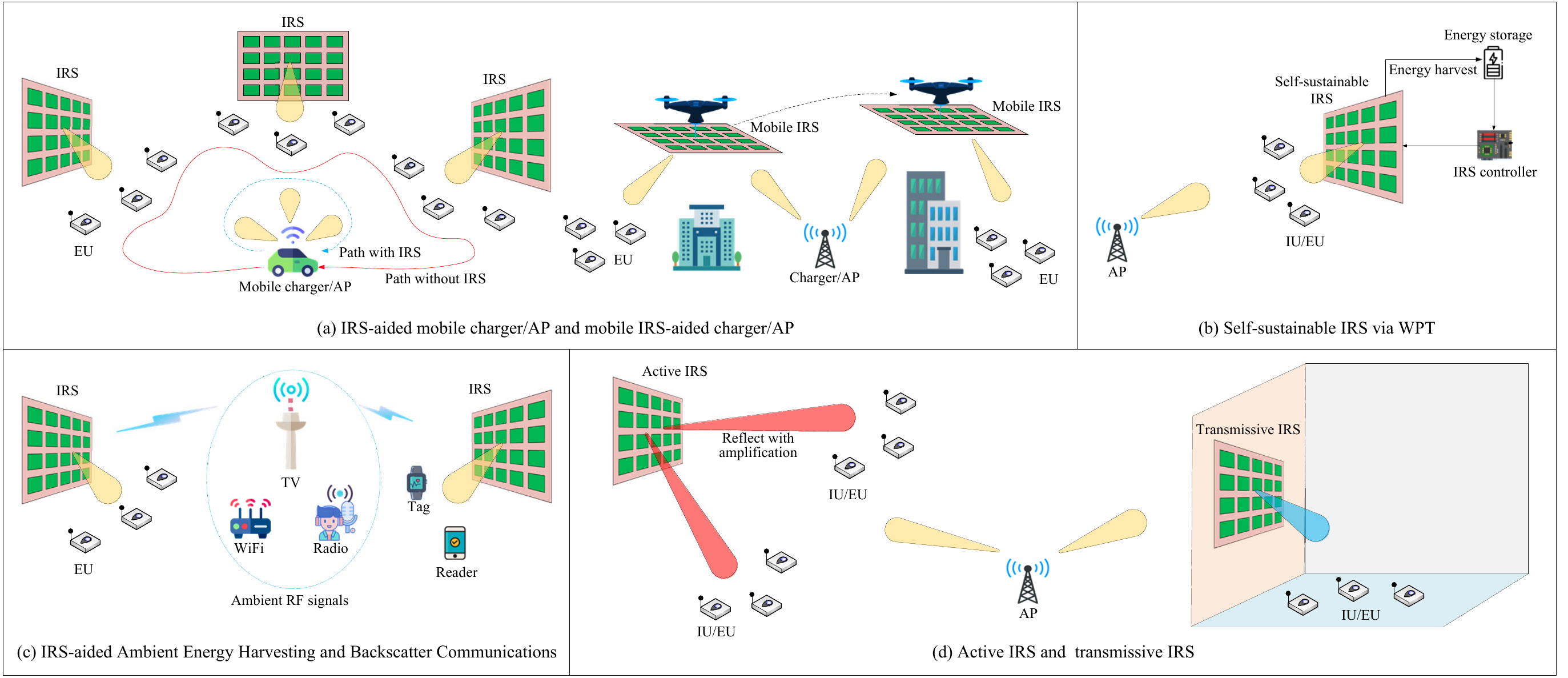}
	\caption{Other application scenarios of IRS in WPT/WIPT networks.}
	\label{future}
	\vspace{-2mm}	
\end{figure}

\subsection{Self-sustainable IRS via WPT} 
Although IRS only passively reflects the impinging signals, the power consumed by its controller for communication and dynamically controlling  its elements cannot be neglected in practical systems, especially when the number of IRS elements is large. When IRS is deployed in places without wired power supply or for applications where it is difficult/inconvenient/costly to replace battery, equipping IRS with RF energy harvesting modules as shown in Fig. \ref{future}(b) is a practical solution to provide IRS additional energy supply which may even achieve its self-sustainable operation.  By treating each IRS element as an EU and depending on the hardware design for separating its two functions of signal reflection and energy harvesting (e.g., antenna/element switching \cite{hu2020sum}, time switching (TS) \cite{zou2020wireless}, power splitting (PS) \cite{lyu2021optimized}), IRS can achieve both functions with flexible tradeoff and use the harvested energy to support its signal reflection.


For each of the above energy harvesting designs, there exists a fundamental tradeoff in maximizing the harvested energy and the time for signal reflection at each IRS. For example, with TS-based energy harvesting/reflection, too little time for energy harvesting at the IRS cannot provide sufficient energy for achieving its optimal reflection performance, while too much time for energy harvesting would result in insufficient time of IRS for aiding the desired WPT/SWIPT/WPCN. It is thus interesting to investigate the time/resource allocation at the IRS to optimize its performance under the new self-sustainable energy consideration in future work.

\subsection{IRS-aided Ambient Energy Harvesting and Backscatter Communications}
So far, we consider dedicated RF sources for WPT and WIPT. While in urban environment, numerous ambient RF signals exist, such as TV, radio, cellular, and WiFi signals, which could be potential  energy sources for energy harvesting. Nevertheless, it is difficult for EUs to efficiently harvest energy from these abundantly available signals due to their randomness in nature. To this end, IRS is highly promising to resolve this problem since the ambient signals can be focused at EUs by exploiting IRS passive beamforming, as shown in Fig. \ref{future}(c), thus significantly improving the energy harvesting efficiency for EUs. However, there is a practical challenge that the CSI on both IRS-reflected and non-IRS-reflected channels from each unknown RF resource to the EU is usually unknown. The energy feedback based beam training scheme proposed in this paper could be a potential solution for overcoming this issue since the passive beam pattern is optimized based on the EUs' energy measurements without the need of estimating CSI explicitly. Note that the ambient RF sources may change over time in terms of signal strength, angle-of-arrival, and so on, thus adaptive IRS reflection is required to track the strongest RF source for energy harvesting.

On the other hand, backscatter communication is a practically appealing technique for enabling low-cost IoT devices such as tags and sensors to send their information to the AP (Reader) without any RF chains \cite{8454398}. In particular, ambient backscatter communication (ABC), in which ambient RF signals are exploited as the free carriers for backscatter communication has received increasing interest recently, as it dispenses with dedicated RF source for sending the carrier signal as in conventional backscatter communication systems. However, a practical issue in backscatter communication arises from the need for canceling the interference at the receiver of  AP/Readrer due to the carrier signal before decoding the desired backscatter device's signal, which becomes more severe in ABC systems due to the unknown carrier signal that is usually modulated with random messages \cite{8368232}.  IRS can be a potential solution to this issue by either enhancing the backscatter signal or suppressing the interference at the AP/Reader receiver via proper signal reflection. However, due to the double passive reflections of both backscatter device and IRS elements, the IRS reflection design and channel acquisition issue become more involved in IRS-aided backscatter communication/ABC systems as compared to the conventional IRS-aided WPT/WPCN systems. Thus, significant research effort is required to tackle these problems in future work.

\subsection{Active/Transmissive Metasurface for WPT/WIPT}

While we have shown that the passive IRS is able to improve the WPT/SWIPT/WPCN performance at low energy and hardware cost, a large number of reflecting elements is practically required due to the product-distance path loss caused by IRS's passive reflection. To overcome this drawback, a new type of IRS, called active IRS, has been recently proposed \cite{long2021active,you2021active}, as shown in Fig. \ref{future}(d). Specifically, by equipping the reflecting elements with negative resistance components such as tunnel diode and negative impedance converter, the active IRS can reflect the incident signal with power amplification at the expense of modestly higher hardware and energy cost. However, the optimal designs for WPT/WIPT with active IRS have remained largely open and a detailed comparison of active IRS versus passive IRS is yet to be investigated under practical setups. On the other hand, the reflection-type metasurface only serves the transmitters and receivers located at the same side, whereas a refraction-type metasurface is useful in enhancing the WPT/WIPT coverage by serving the transmitters and receivers located at its opposite sides. As such, the research on WPT and WIPT aided by transmissive IRS is another important topic to pursue in future work.



\section{Conclusions}

In this article, we have provided a tutorial overview on the main communication and signal processing issues and their potential solutions  for IRS-aided WPT as well as IRS-aided SWIPT and WPCN. In particular, IRS reflection design, channel acquisition, and resource allocation optimization were thoroughly studied, by taking into account the unique performance metrics and hardware constraints of EUs. It was shown that IRS is able to significantly improve the efficiency of WPT, which also benefits the rate-energy tradeoff in SWIPT and the wireless-powered communication performance in WPCN.  Furthermore, we discussed open issues and other related topics that are worthy of further research and investigation. It is hoped that this paper would provide useful guidance for future research into this emerging and promising area, to  unlock the full potential of IRS for achieving cost-effective and sustainable wireless networks in the future.

%
%
%
%
%

\bibliographystyle{IEEEtran}
\bibliography{IEEEabrv,mybib}
\end{document}